\documentclass[12]{article}
\usepackage{amsmath}
\usepackage{color}

\definecolor{red}{rgb}{1,0,0}
\definecolor{blue}{rgb}{0,0,1}
\definecolor{green}{rgb}{0,1,0}
\definecolor{black}{rgb}{0,0,0}
\definecolor{yellow}{rgb}{1,1,0}
\definecolor{mdwblue}{rgb}{0.2,0.2,0.6}
\definecolor{gray}{rgb}{0.7,0.7,0.7}
\definecolor{darkgreen}{rgb}{0.2,0.7,0.2}
\begin{document}
\title{Generalized Gravity I : Kinematical Setting and reformalizing Quantum Field Theory.}
 \author{Johan Noldus\footnote{johan.noldus@gmail.com}} \maketitle
\begin{abstract}
The first part of this work deals with the development of a
natural differential calculus on non-commutative
manifolds.  The second part extends the covariance and equivalence principle as well
studies its kinematical consequences such as the arising of gauge theory.  Furthermore, a
 manifestly causal and covariant
 formulation of quantum field theory is presented which surpasses the usual Hamiltonian
  and path
 integral construction.  A particular representation of this theory on the kinematical
 structure developed in section three is moreover given.
\end{abstract}
\section{Introduction}
In this paper, we follow the axiomatic approach towards the derivation of physical laws; that is our basic question is ``if the universe were
a self computing entity, then what are its rules?''.  The idea is to find a set of (almost) evident laws in order to arrive at a
unique mathematical structure determining the dynamical equations;
the latter being the message conveyed by general covariance
and a form of the equivalence principle.  It goes without saying
that the resulting equations need to be brought into correspondence
with the models constructed from direct observation.  In the same vein, it would be great progress if one could find an axiomatic base for quantum mechanics given that the current quantum recipe proves to be troublesome in conjunction with gravitational physics.  So, where do we look for this
theory?  The only theoretical principles in physics which in my
opinion should be taken seriously are the principle of general
covariance as well as the equivalence principle, but the latter
should not be applied to gravity only.  Assuming furthermore the
continuum hypothesis, it is clear that the first task to perform is
to develop non-commutative calculus.  The latter constitutes the
first part of this work in which a general theory of non-commutative
manifolds, tensor and differential calculus is developed.  Although the specific manifold structure
we arrive at in section three allows for a more economic geometrical treatment, it is nevertheless
a good excericise to go through this idea.  When doing non-commutative geometry, one could on one hand
try to look for a representation in terms of abelian coordinates but with a second deformed product
structure.  The particular abelian realization of the Moyal-$\star$ product then serves to transport
the standard abelian differential calculus and allows one to study non commutative deformations of Gravity.
One might, on the other hand, take the complete opposite point of view and start from non commutative algebra's
 without bothering about abelian representations.  Certainly, this comes at a cost if no algebraic properties are
being preserved, but the latter is manageble nevertheless.  Ultimately, we shall work with variables having special algebraic properties
and we shall show how the resulting calculus can be embedded into our general framework.  It is nevertheless very useful
to dispose of both views and the tension between both calculi resides around the validity of the Leibniz rule.  But again,
sometimes one should look around like a bird prior to indulging into more practical matters and indeed, the abstract theory, apart from being more compact and intuitive
gives all the necessary tools.  Our definitions imply the mathematical curiosity which states that,
 in the abelian limit, space time gets one dimensional as a non
commutative manifold and is therefore flat.  Given these results, we
look for natural mathematical demands in order to select the correct
manifold and from thereon develop a gravitational theory. The latter
constitutes the second part of this work and it turns out in section
$3.1$ that merely five natural axioms fix the algebra in the correct
way so that (a) the number of Standard Model generations is derived as well as the
number of Dirac spinors per generation.  Given that bi-quaternions have an important geometrical
role to fulfill in the kinematical setting at hand, we spent the entire section $3.4$ at studying free
Fermi field theory in the quaternion formalism.  This turns out to be a very delicate task
 which suggests a drastic reformulation of gauge theory which we
accomplish in subsection $3.5$.  Not only do we reformulate gauge theory, but we have answered by then
 the question ''Why gauge theory?''.  Indeed, albeit the usual question is ''Why the standard
  model?'', the former
 question has more serious implications for the kind of mathematics required to describe
  nature than the latter.  The existence of gravity is god given by insisting upon general covariance
  but which \emph{natural} symmetries do give rise to gauge forces?  The answer suggested in
  this paper is that for general algebra valued coordinates, it is possible for the equations of nature
  to care about \emph{real} parts of them; that is involutions - which are nothing but idempotent algebra
   anti-automorphisms- appear in almost every physical law.  The gauge principle then emerges from the observation
   that physical laws should be invariant under suitable rotations of these involutions.  Actually, this
   viewpoint allows us to \emph{derive} some fundamental
 constants of nature and in section $3.5$, we present some details for the $U(1)$ gauge group.
   In section $3.6$, we re-axiomatize
free quantum field theory in the quaternion formalism, show how the latter has a geometrical
representation on our spacetime and comment upon the inclusion of interactions.
While doing so, we learn that quantum really means ``infinitesimal atomization'' of the
 abelian component of space time (with infinitesimal penetrations in the non commutative
 structure) in close resemblance to the derivation of statistical mechanics from classical
  field theory.  In section $3.7$, it is shown that our space time manifold naturally gives rise to a bundle structure and the latter coincides with the unbroken symmetries of the standard model, that is $U(1) \otimes SU(3)$.  Moreover, we show that the charges associated to the electromagnetic interactions all come out correctly.  However, I do at this stage not know yet how to include the weak interactions as well as all correct coupling terms with the quarks and Higgs.  Evidently, this approach has some further difficulties which need to be resolved such as stability issues related to the possibility for matter fields to penetrate the non commutative components. \\* \\
Given the size of the above work, I have opted for using this natural breaking point and postpone the issue of dynamics to the second paper.  The latter shall have a similar structure to this one: first non commutative integral calculus and co-homology theory shall be worked out, second we formulate the dynamical laws and work out the correspondence to gravity and the standard model on one hand and calculate some explicit solutions on the other. \\* \\
The author apologizes in advance for totally omitting a literature list for the simple reason that no sources were used for this work and therefore, the particular relevance of a citation and the absence of others might be greater source of confusion.
\section{Non abelian geometry.}
\subsection{Non abelian manifolds.} What follows below is not restricted to finite Von Neumann algebras and where necessary, the relevant technical details can be easily filled in. \\* \\ Let $\mathcal{A}$ be a finite
unital real or complex Von Neumann algebra, that is $\mathcal{A}$ as a vector space over $R$ or $C$
has finite dimension and a conjugation and trace functional $tr$ are
defined upon it.  Henceforth, we use $K$ as a shorthand for either $R$ or $C$.  Some mathematical results are
\begin{itemize}
\item $\mathcal{A}$ has a Pauli Cartan basis over $K$, that is a basis
$e_{\mu}$ of self adjoint elements such that $tr(e_{\mu} e_{\nu}) =
\delta_{\mu \nu}$ where $\mu,\nu: 1 \ldots N$.
\item $\mathcal{A}$ is finitely generated by $z^{\alpha}$, $\alpha :
1 \ldots M \leq N$ over a subalgebra $\mathcal{B}$ containing $K$, meaning that any element of $\mathcal{A}$ belongs
to the polynomial algebra $\mathcal{B}(1,z^{\alpha})$ consisting of
monomials in $1,z^{\alpha}$ with coefficients in $\mathcal{B}$.
\end{itemize}
\underline{Remark} \\ Assuming $\mathcal{B} = C$, the number of
 generators is equal to the number of basis elements minus one if
 $\mathcal{A}$ is generated by idempotent or nilpotent elements only\footnote{A proof of this statement is easy,
 suppose $a$ is a generator independent from unity; since
 the number of generators equals the number of basis elements minus one $a^2 =
 \alpha a + \beta$.  Replacing $a$ by $(a + \epsilon 1)$ where $\epsilon$ is a root of $x^2 + \alpha x - \beta = 0$, results in $a^2 = (\alpha + 2 \epsilon) a$.  If the latter equals zero, then $a$ is a nilpotent generator, otherwise one can scale
 $a \neq 1$ to an idempotent element.}. \\* \\ First, let us critically examine the definition of an ordinary manifold, say $R^{8}$; the latter is topologically equivalent to $H \oplus H$, with $H$ the algebra of real quaternions which is generated by two elements as an algebra over $R$.  Therefore, the question which naturally arises is why we speak about the real polynomial algebra in $8$ commuting variables instead of the real polynomial algebra in two quaternion variables?  Hence, the definition of a manifold should incorporate the algebraic properties of the variables in which one wishes to calculate.  It seems furthermore a natural idea that the algebra determines the dimension of the manifold as we shall see now.  Denote by $\mathcal{S} \subset \mathcal{A}^{M}$ the
subset of generating $M$-tuples (in the natural topology) which we assume\footnote{It would be nice to have some general structural insight into this issue.} to be open, and let
$x^{\alpha}$ be an $M$-tuple of variables taking values in
$\mathcal{S}$.  The $x^{\alpha}$ are to be regarded as the
coordinates of our non commutative manifold $\mathcal{S}$.
\newtheorem{Definition}{Definition}
\begin{Definition}
Given a finite unital Von Neumann algebra $\mathcal{A}$ and subalgebra $\mathcal{B}$, containing $K$, with
associated space of generators $\mathcal{S}$.  An
$(\mathcal{A}, \mathcal{B})$-manifold $\mathcal{M}$ is by definition a topological
space which is locally homeomorphic to $\mathcal{S}$ and locally carries the function algebra $\mathcal{A}(1,x^{\alpha})$ where the $M$-tuple $x^{\alpha}$ is $\mathcal{S}$-valued.
\end{Definition}
More in detail, let $p \in \mathcal{M}$, then there exists a pair
$(\mathcal{O},\psi)$ where $p\in \mathcal{O}$ and $\psi :
\mathcal{O} \rightarrow \mathcal{S}$ an injective mapping.  Given
two such pairs $(\mathcal{O}_1,\psi_1)$ and
$(\mathcal{O}_2,\psi_2)$, then $\mathcal{M}$ is a $C^{0}$ manifold
if and only if $\psi_2 \circ \psi_1^{-1} : \psi_1(\mathcal{O}_1 \cap
\mathcal{O}_2) \rightarrow \psi_2(\mathcal{O}_1 \cap \mathcal{O}_2)$
is a homeomorphism.  $R^n$ as a topological space is a
$(R^n,R)$ manifold and $\mathcal{S}$ consists of one
generator only.
\subsection{Tensor calculus.} Presenting a ``new'' theory is
always a delicate choice between giving abstract rules and computing
some examples on one hand and trying to build it while making
mistakes on the other.  As a result, I have opted for subdividing
the definition according to different natural questions with a
unified treatment following at the end.  Our aim is to generalize
Taylor's rule, the difference with the abelian case being that the
infinitesimals cannot a priori be shifted to the left or right so
that position in a word must be remembered, this is the task of the
symbol $\omega$.  On the other hand, when we evaluate the
differential on an algebra element $a$, an operation which shifts
$a$ to the position of $\omega$ and annihilates the latter is needed. This
is the role of the associative product $\star$.  By convention,
members of the $\mathcal{S}$-valued $M$-tuple of variables are
denoted by $x^{\alpha}$ and constant algebra elements by Latin
letters $a,b,c \ldots$.
\begin{Definition}
$V^{(1)}(\omega)$ is the finite dimensional vectorspace over $K$
formed by linear combinations of elements $a\omega b$ where $a,b \in
\mathcal{A}$.  It can be made into a unital, associative algebra
$(V^{(1)}(\omega), \star)$ by defining $(a \omega b) \star (c \omega d) = ca \omega bd$.  More generally $(x)\star a = a(x) \star$ and $(\omega)\star x = x = (x)\star \omega$ where $x \in
V^{(1)}(\omega)$. $\omega$ is a unit and by convention $x \star y$ =
$(x)\star y$.
\end{Definition}
By definition, one also has that $\left[ a , (b)\star \right] = 0$
and $a = (a)\star \omega$.  Since traces can be taken of algebra
elements and derivatives commute with traces, we should allow for
words containing formal traces of $\omega$ satisfying $$(a)\star tr(\omega) = tr(a).$$  Actually, the latter is part
of a general procedure which we may call trace completion.
\begin{Definition}
Let $\mathcal{A}$ be an associative algebra and $tr$ a formal trace functional; then the trace completed algebra $\mathcal{A}^{tr}$
is generated by $a$ and $tr(b)$ where $a,b \in \mathcal{A}$.
\end{Definition}
An obvious property in $V^{(1)}(\omega)$ is $\ldots (b a)\star =
\ldots (b)\star (\omega
 a)\star$ and by definition $\star$ commutes with the trace.  In the
 sequel $W^{(1)}(\omega)$ is the trace closure of the unital $\star$ algebra
 generated by elements of the form $x\omega y$ with $x,y \in
 \mathcal{A}(x^{\alpha})^{tr}$.
\begin{Definition}
Note by $\mathcal{A}(x^{\alpha})$ the algebra of polynomials in the
variables $x^{\alpha}$ and constants $a,b,c \ldots \in \mathcal{A}$.
The partial derivatives $\partial_{x^{\alpha}} :
\mathcal{A}(x^{\alpha})^{tr} \rightarrow W^{(1)}(\omega)$ are
defined by $\partial_{x^{\alpha}} x^{\beta} =
\delta^{\beta}_{\alpha} \omega + x^{\beta}
\partial_{x^{\alpha}}$, $\left[ \partial_{x^{\alpha}} , a \right] =
0$ and $\partial_{x^{\alpha}}$ commutes with the trace.
\end{Definition}
Clearly, the Leibnitz rule is satisfied and the chain rule has a
substitute.  Let $x'^{\alpha}(x^{\beta})$ and $g$ be a
differentiable function, then
$$ \partial_{x^{\alpha}}
(g(x'^{\beta}(x^{\gamma}))) =  \left(
 \partial_{x^{\alpha}}
(x'^{\mu}(x^{\gamma})) \right)\star \partial_{x'^{\mu}}
(g(x'^{\beta}(x^{\gamma})))$$ Therefore \begin{eqnarray*}
\delta^{\alpha}_{\gamma}\omega & = & \left( \frac{\partial
x'^{\beta}}{\partial x^{\gamma}} \right)\star \frac{\partial
x^{\alpha}}{\partial x'^{\beta}}
\end{eqnarray*} modulo derivatives of the Cayley-Hamilton identities. \\*
\\
\underline{The algebra $(V^{(1)}(\omega)^{tr},\star)$} \\* \\
Let $e_{\mu}$ be a Pauli Cartan basis of $\mathcal{A}$, then it is
easily seen that any element $x$ of $V^{(1)}(\omega)^{tr}$ can be
uniquely decomposed as $x = x^{\mu} tr(\omega e_{\mu})$ where
$x^{\mu} \in \mathcal{A}$. We prove that $x\star y= \omega$ if and
only if $y \star x = \omega$.  Define the matrix $G^{\mu}_{\nu}(x)
:=
\begin{array}{c}
 tr(e^{\mu}x_{\nu})  \\
\end{array}$ where lowering and raising of the indices occurs with
$\delta_{\mu \nu}$ and $\delta^{\mu \nu}$ respectively.  Then, $x =
e_{\mu} G^{\mu}_{\nu}(x) tr(\omega \, e^{\nu})$ and since
$tr(e_{\mu} e^{\nu}) = \delta^{\nu}_{\mu}$ one arrives at
$$y \star x = e_{\mu} G^{\mu}_{\alpha}(x) G^{\alpha}_{\nu} (y) tr(\omega \, e^{\nu})$$ which proves our assertion since $G(x)G(y) = 1$ if and only
$G(y)G(x)$ is.  The result for matrix algebra's over this algebra
obviously follows since
$$e_{\mu} G^{\mu}_{\nu} (x_{i}^{j}) tr(\omega e^{\nu}) \star  e_{\kappa} G^{\kappa}_{\lambda}(y_{j}^{k}) tr(\omega \, e^{\lambda})  = $$
$$ e_{\mu} G^{\mu}_{\kappa}(y_{j}^{k})G^{\kappa}_{\nu}(x_{i}^{j}) tr(\omega \, e^{\nu}) = \omega \, \delta_{i}^{k}$$ if and only if $G^{\mu}_{\kappa}(y^{k}_{j})G^{\kappa}_{\nu}(x^{j}_{i})
= \delta^{\mu}_{\nu} \delta^{k}_{i}$.
Using the equality
$$\frac{\partial x'^{\beta}}{\partial x^{\alpha}} = e_{\mu} \left(
tr\left( (e_{\nu})\star \frac{\partial x'^{\beta}}{\partial
x^{\alpha}} e^{\mu} \right) \right)tr(\omega e^{\nu})$$ where the
complex matrix
$$C^{\beta \mu}_{\,\, \alpha \nu}(x';x) = \left(
tr\left( (e_{\nu})\star \frac{\partial x'^{\beta}}{\partial
x^{\alpha}} e^{\mu} \right) \right)$$ is invertible, the previous
equality is equivalent to
$$C^{\alpha \mu}_{\, \beta \nu} (x';x) C^{\beta \nu}_{\, \gamma
\xi}(x;x') = \delta^{\alpha}_{\gamma} \delta^{\mu}_{\xi}.$$  Using
algebra, many expressions of the form $(a)\star x$ can be simplified
due to the Cayley Hamilton identities $CH(z) = 0$.  It is important
to understand that these identities do not play any part in our
differential calculus, but that the latter is consistent with them
in the following sense:
$$(a)\star \partial_{x^{\alpha}} CH(z) = 0$$ where $z \in
\mathcal{A}(x^{\beta})$.  Hence, it is important to realize that
two expressions $x,y \in W^{(1)}(\omega)$ are equal $x = y$ if and
only if $a\star \, x = a\star \, y$ for any $a \in \mathcal{A}$. \\*
\\  The transformation rules for derivatives are given by
$$\partial_{x^{\alpha}} = C^{\beta}_{\alpha}(x';x) \star
\partial_{x'^{\beta}}.$$  Vector fields are defined as $$V = V^{\alpha}\star \partial_{x^{\alpha}}$$ where
$V^{\alpha} \in W^{(1)}(\omega)$ and $V'^{\alpha} = V^{\beta}\star
C^{\alpha}_{\beta}(x';x)$.  One defines a dual basis $dx^{\alpha}$
by requiring that $$\partial_{x^{\beta}} \star dx^{\alpha} =
\delta^{\alpha}_{\beta}\omega$$  Obviously, $dx^{\alpha}$ transforms
as $dx^{\alpha} = dx'^{\beta} \star C^{\alpha}_{\beta}(x;x')$ since
$$\partial_{x^{\beta}} \star dx^{\alpha} = C^{\gamma}_{\beta}(x';x)
\star \partial_{x'^{\gamma}} \star dx'^{\kappa} \star
C^{\alpha}_{\kappa} (x;x') = C^{\gamma}_{\beta}(x';x) \star \omega
\star C^{\alpha}_{\gamma} (x;x') = \delta^{\alpha}_{\beta} \omega.$$
Dual fields are defined as $dx^{\alpha} \star W_{\alpha}$ where
$W'_{\alpha} = C^{\beta}_{\alpha}(x;x') \star W_{\beta}$ and
therefore contractions $V^{\alpha} \star W_{\alpha}$ are scalars
belonging to $W^{(1)}(\omega)$ since obviously $W_{\alpha} \in
W^{(1)}(\omega)$.  \\* \\ \underline{Remark} \\ It is possible to
define $V^{\alpha} \in \mathcal{A}(x^{\beta})^{tr}$ so that
$V^{\alpha} \star W_{\alpha} \in \mathcal{A}(x^{\beta})^{tr}$.
Notice that in case $V^{\alpha} \in W^{(1)}(\omega)$, the invariance
property for $V^{\alpha} \star W_{\alpha}$ has to be understood
modulo contractions with $(a)\star$ in order to eliminate
derivatives of Cayley-Hamilton identities.  In the latter case,
where $V^{\alpha} \in \mathcal{A}(x^{\beta})^{tr}$, these vanish
automatically. \\* \\
Prior to developing higher differential calculus, we should finally
say a word about generalized diffeomorphisms. \\*
\\ \underline{Note on coordinate transformations} \\  We might have
started this text by using the natural fibration induced by the
trace part of the generators; indeed we could have spoken about
bosonic and fermionic derivatives. It is natural to look for
embeddings of the diffeomorphism algebra of the trace variables into the
larger algebra of local diffeomorphisms in the generators of
$\mathcal{A}$. Given a polynomial local diffeomorphism $\phi$ in the
trace variables; $\phi^{\alpha}$ can always be written as:
$$\phi^{\alpha} = \psi^{\alpha} + tr(x^{\alpha})\varphi^{\alpha}$$
where $\psi^{\alpha}$ is independent of $tr(x^{\alpha})$.  Hence,
the obvious candidate for a local algebra diffeomorphism is
$$x'^{\alpha} = \frac{1}{tr(1)} \psi^{\alpha} + x^{\alpha}\varphi^{\alpha}.$$
Clearly, the inverse and implicit function theorem can be applied to
diffeomorphisms of the above kind using the generalized Jacobian
$C^{\alpha \mu}_{\,\beta \nu}(x';x)$. \\* \\  Applying an operator
of the kind $z_1 \star
\partial_{x^{\alpha}}$ twice on a $\mathcal{A}$-valued function
$\Psi \in \mathcal{A}(x^{\beta})^{tr}$ results in
$$z_1 \star \partial_{x^{\alpha}} \left( (z_2)\star \partial_{x^{\beta}} \Psi \right) = (z_1)\star (\partial_{x^{\alpha}}z_2)\star
\partial_{x^{\beta}} \Psi + (z_1)\star (z_2)\star
\partial^{(2)}_{x^{\alpha}} \partial_{x^{\beta}} \Psi$$ where $z_j \in
\mathcal{A}(x^{\alpha})^{tr}$.  At least, an expression of this kind
with higher derivatives is desirable and one is left with the
problem of finding efficient and insightful algebra language
realizing this.  For this, we introduce an infinity of new elements
$\omega^{(k)}$ satisfying the following algebra
\begin{itemize}
\item $\omega^{(k)} \star \omega^{(l)} = \omega^{(l-1)} \left( \omega^{(k-1)}
\right)\star$ for $k \geq l > 1$,
\item $(a)\star \, \omega^{(k)} = \omega^{(k-1)} \, (a)\star$ for
all $k > 1$,
\item $\omega^{(k)} \star \omega
=  \omega^{(k)}$,
\item $(\omega \, \omega^{(k)})
\star \omega^{(l)} = \omega^{(l)} (\omega \, \omega^{(k-1)})\,
\star$ for all $k \geq l > 1$,
\item $(\omega \, a)\star \, \omega^{(k)}
= \omega^{(k)} \, (\omega \,a)\star$ for all $k > 1$,
\item $(\omega \, \omega^{(k)})\star \omega
=  \omega \, \omega^{(k)} = \omega \star (\omega \, \omega^{(k)})$
for all $k>1$,
\item $\omega^{(k)} \star \omega^{(l)} = \omega^{(l)}
\left(\omega^{(k)} \right) \star$ for $1 \leq k < l$,
\item $(\omega \, \omega^{(k)})\star \omega^{(l)} = \omega^{(l+1)}
\left(\omega \omega^{(k)} \right)\star$ for $1 \leq k < l$.
\end{itemize}
and $(\textrm{anything})\star$ commutes with $\mathcal{A}$-valued
elements.  Let $\mathcal{A}(\{\omega^{(k)}\}_{k
> 0})$ be the $\mathcal{A}$-module of words containing at least one of the
generators $\omega^{(l)}$ and define partial differential operators
by
$$\partial^{(k)}_{x^{\alpha}} x^{\beta}  = \delta^{\beta}_{\alpha} \omega^{(k)} + x^{\beta}
\partial^{(k)}_{x^{\alpha}}$$ where $$\partial^{(k)}_{x^{\alpha}}
\omega^{(l)} = \omega^{(l)} \partial^{(k-1)}_{x^{\alpha}}$$ for $k >
l \geq 1$ and $$ \partial^{(k)}_{x^{\alpha}} \star = \star
\partial^{(k+1)}_{x^{\alpha}}$$ for all $k>0$.  It is
sufficient to verify
$$a \star
\partial_{x^{\alpha}} \left( (b)\star \partial_{x^{\beta}} \Psi
\right) = (a)\star (b)\star
\partial^{(2)}_{x_{\alpha}} \partial_{x^{\beta}} \Psi$$ on a
monomial $\Psi(x^{\beta}) = x^{\alpha_1} a_1 x^{\alpha_2} a_2 \ldots
a_{n-1}x^{\alpha_n}$ since all derivatives commute with the trace
and the Leibnitz rule applies.  The left hand side of this
expression is a sum over all couples $(i,j)$, $i \neq j$, where in
the corresponding term $x^{\alpha_i}$ is replaced by
$\delta^{\alpha_i}_{\alpha} a$ and $x^{\alpha_j}$ by
$\delta^{\alpha_j}_{\beta} b$ in the expression $x^{\alpha_1} a_1
x^{\alpha_2} a_2 \ldots a_{n-1}x^{\alpha_n}$.  The right hand side
is given by
$$a \star (b)\star \left(\sum_{i < j} x^{\alpha_1} a_1 \ldots
\delta^{\alpha_i}_{\alpha} \omega^{(2)} a_i \ldots
\delta^{\alpha_j}_{\beta} \omega a_j \ldots a_{n-1}x^{\alpha_n} +
\sum_{i > j} x^{\alpha_1} a_1 \ldots \delta^{\alpha_j}_{\beta}
\omega a_j \ldots \delta^{\alpha_i}_{\alpha} \omega a_i \ldots
a_{n-1}x^{\alpha_n} \right)$$ and this can be reduced to
$$\sum_{i < j} x^{\alpha_1} a_1 \ldots
\delta^{\alpha_i}_{\alpha} a a_i \ldots \delta^{\alpha_j}_{\beta} b
a_j \ldots a_{n-1}x^{\alpha_n} + \sum_{i > j} x^{\alpha_1} a_1
\ldots \delta^{\alpha_j}_{\beta} b a_j \ldots
\delta^{\alpha_i}_{\alpha} a a_i \ldots a_{n-1}x^{\alpha_n}$$ which
proves our statement. The more general case can be proven by
noticing that for any $z \in \mathcal{A}(\{\omega^{(k)}\}_{k \geq
l})$ and $l > 1$, $z \star \omega^{(l)} = \omega^{(l-1)} (z')\star$
where $z'$ is obtained from $z$ by replacing each $\omega^{(k)}$ by
$\omega^{(k-1)}$ for $k \geq l$. Indeed any word of the form $a_1
\omega^{(k_1)} \ldots a_{n} \omega^{(k_n)}a_{n+1} \star$ can be
decomposed as
$$(a_1)\star
(\omega \, \omega^{(k_1)})\star \ldots (\omega \, a_{n})\star
(\omega \, \omega^{(k_n)})\star (\omega \, a_{n+1})\star.$$  Prior
to proceeding with the reconstruction of Taylor's rule and higher
tensor algebra, it is instructive to
apply this calculus to the algebra $R^n$. \\* \\
\underline{Example} \\* \\ $R^{n}$ can be endowed with the natural
product $$(a_1, \ldots , a_n)(b_1, \ldots , b_n) = (a_1 b_1 , \ldots
,a_n b_n)$$ and has $n$ commuting basis vectors and one generator.
Considering $R^n$ as a $(R^n,R)$ manifold, one disposes of one
variable $x \in \mathcal{S}$ and the relevant function space is
$R^n(x)^{tr}$ where $tr(a_1, \ldots , a_n) = \sum_{j=1}^n a_j$.  The
latter coincides with the usual function space: $x^j$ may be
identified with $(0, \ldots ,0,1_j,0,\ldots,0)x = (0, \ldots , 0,
x_j , 0 , \ldots , 0)$.  In order to swap $x_j$ from the $j$'th to
the $k$'th index it suffices to consider combinations such as
$$tr(x^j)(0, \ldots, 0 , 1_j , 0 ,\ldots , 0, 1_k ,0, \ldots , 0) -
x^j$$ where in this case $j<k$.  As such, all generators
$(x_j)_{j=1}^{n}$ are freely available on each index.  The natural
differential on $R^n(x)^{tr}$ is given by $\partial_x$ and it is
easy to see that $(0, \ldots ,0,1_j,0, \ldots,0)\star
\partial_x = \partial_{x_j}$; in particular its action on monomials $x^m$
is given by
$$(a) \star
\partial_x x^m = (a)\star m x^{m-1} \omega = mx^{m-1}a$$ where
$\omega$ may be shifted trough since $R^n$ is abelian.  The last
expression coincides with $$\left( \sum_{j=1}^{n}a_j
\partial_{x_j} \right) \left(
\begin{array}{c}
  x_1^m \\
  \vdots \\
  x_n^m \\
\end{array}
\right).
$$ \\ \underline{Note} \\ Obviously, $\partial^{(2)}_{x^{\alpha}} \partial_{x^{\beta}} \neq  \partial^{(2)}_{x^{\beta}}
\partial_{x^{\alpha}}$ for $\alpha \neq \beta$ but $$(a)\star (b)\star \partial^{(2)}_{x^{\alpha}} \partial_{x^{\beta}} =  (b)\star (a)\star \partial^{(2)}_{x^{\beta}}
\partial_{x^{\alpha}}.$$
Taylor's rule is given by
 $$\Psi(x^{\alpha} + \theta^{\alpha}(x^{\beta})) - \Psi(x^{\alpha})
 = \sum_{n=1}^{\infty} \frac{1}{n!}\sum_{(\beta_1, \ldots, \beta_n) \in M^n} (\theta^{\beta_n})\star
 \ldots (\theta^{\beta_1})\star \partial^{(n)}_{x^{\beta_{n}}}
 \partial^{(n-1)}_{x^{\beta_{n-1}}} \ldots \partial_{x^{\beta_1}}
 \Psi(x^{\alpha})$$ and the proof is left as an easy exercise to
 the reader.  Note that $W^{(1)}(\omega)$ is much
 bigger than $\partial_{x^{\alpha}}\left( \mathcal{A}(x^{\alpha})^{tr}
 \right)$ for any $\alpha$ because of the non commutativity; indeed, $\sum_{\alpha =1}^{M} \omega \, tr(x^{\alpha})$ is such an
 element.  Therefore, we define the $\mathcal{A}$-module $Z^{k}(\{ \omega^{(l)} \}_{l \leq
 k})$ as $$Z^{k}(\{ \omega^{(l)} \}_{l \leq
 k}) = \{ \partial^{(k)}_{x^{\alpha_{k}}}
 \partial^{(k-1)}_{x^{\alpha_{k-1}}} \ldots
 \partial_{x^{\alpha_{1}}} \Psi \, | \, (\alpha_{1} , \ldots ,
 \alpha_{k}) \in M^k \textrm{ and } \Psi \in
 \mathcal{A}(x^{\alpha})^{tr} \}.$$  We shall now further
 characterize the $\mathcal{A}$-modules $Z^{k}$; $Z^{1}(\omega) \subset
 W^{(1)}(\omega)$ and any element in
 $Z^{2}(\omega,\omega^{(2)})$ is a linear combination of elements of the form $z_1 \omega z_2 \omega
 z_3$, $z_1 \omega^{(2)} z_2 \omega
 z_3$, $tr(z_1 \omega)z_2 \omega z_3$, $tr(z_1 \omega^{(2)})z_2 \omega z_3$, $z_1 \omega z_2 tr(\omega z_3)$, $z_1 \omega^{(2)} z_2 tr(\omega z_3)$ and
traces thereof where $z_i \in \mathcal{A}(x^{\alpha})^{tr}$.  The
latter elements are algebraically special since they span the
subspace of elements $w \in (\mathcal{A}(x^{\alpha})^{tr})(\omega,
\omega^{(2)})^{tr}$ satisfying $(z_1)\star (z_2)\star w \in
\mathcal{A}(x^{\alpha})^{tr}$ for all $z_i \in
\mathcal{A}(x^{\alpha})^{tr}$.  In general, denote by
$W^{(k)}(\{\omega^{(l)}\}_{l \leq k} )$ the
$\mathcal{A}(x^{\alpha})^{tr}$ submodule of
$(\mathcal{A}(x^{\alpha})^{tr})(\{\omega^{(l)}\}_{l \leq k})^{tr}$
consisting of those elements $w^{k}$ such that $$(z_1)\star \ldots
(z_{k})\star w^{k} \in \mathcal{A}(x^{\alpha})^{tr}$$ for all $z_{j}
\in \mathcal{A}(x^{\alpha})^{tr}$.  As indicated previously $W^{(k)}(\{\omega^{(l)}\}_{l \leq k} )$ is
spanned by elements $w^{\sigma}$ of the form $z_1 \omega^{(r_1)} z_2
\ldots z_k \omega^{(r_k)} z_{k+1}$ where the series $(r_1,\ldots ,
r_k)$ corresponds to a permutation $\sigma \in S_k$. Indeed, given a
permutation $\sigma$, one constructs a series $(r_1,\ldots ,r_k)$
such that after evaluation $$(a_{k})\star (a_{k-1})\star \ldots
(a_{1})\star w^{\sigma} = z_1 a_{\sigma(1)} z_2 a_{\sigma(2)}
\ldots z_{(k)} a_{\sigma(k)} z_{(k+1)}.$$  The latter is build as
follows : (a) $r_{\sigma(1)} = 1$ (b) if $s_{\sigma(p)}$ is the
number of $\sigma(q)$ where $q < p$ such that $\sigma(p) <
\sigma(q)$, then $r_{\sigma(p)} = s_{\sigma(p)} + 1$.  Taking traces
of subexpressions of $w^{\sigma}$ does not change the above ordering
property and the $(\mathcal{A}(x^{\alpha}))^{tr}$ submodule of
elements with the $\sigma$ property is denoted by
$W^{\sigma_k}(\{\omega^{(l)}\}_{l \leq k} )$.  We show that for any
$w \in W^{(k)}(\{\omega^{(l)}\}_{l \leq k} )$ the following holds
$$\underbrace{(\omega^{(k)})\star \ldots (\omega^{(k)})\star }_{k \textrm{ times}} w =
w.$$  Clearly, it is sufficient of verify this on an element of the
kind $w^{\sigma}$; the first $\omega^{(k)}$ from the right shifts
$\sigma(1) - 1$ times through elements of the kind $\omega^{(l)}$,
$l \geq 2$ resulting in a $\omega^{k - \sigma(1) + 1}$ which is to
be expected since $k - \sigma(1)$ elements should still shift
through.  Given the $p$'th element from the right, $p > 1$, then
shifting $\omega^{(k)}$ trough results in a $\omega^{(k - \sigma(p)
+ 1)}$ while $k - \sigma(p) - s_{\sigma(p)}$ elements still have to
go through.  Hence, $\omega^{( s_{\sigma(p)} + 1)} =
\omega^{(r_{\sigma(p)})}$ remains which we had to prove.   Also, $(\omega^{(k)}) \star w \in W^{(k)}(\{\omega^{(l)}\}_{l \leq k} )$
meaning that $\omega^{(k)}$ is a $k$'th root of a ``left unity'' with respect to $\star$ in the submodule  $W^{(k)}(\{\omega^{(l)}\}_{l \leq k} )$.
 Direct verification of the definition implies that
$$W^{(k)}(\{\omega^{(l)}\}_{l \leq k} ) \star W^{(s)}(\{\omega^{(l)}\}_{l \leq s} )  \subset W^{(k+s-1)}(\{\omega^{(l)}\}_{l \leq k+s-1} ).$$
Moreover, it immediately follows that the product $w^{\sigma} \star
w^{\varsigma}$ with $\sigma \in S_k$, $\varsigma \in S_s$ and
corresponding sequences $(r_{1} \ldots , r_{k})$, $(t_1, \ldots ,
t_s)$, corresponds to a permutation $\sigma \star \varsigma$
determined by the sequence $$(t_1 + k - 1, \ldots t_{\varsigma(1) -
1} + k - 1, r_1 , r_2 , \ldots , r_k , t_{\varsigma(1) + 1}, \ldots
t_{\varsigma(s)}).$$ \underline{Remark} \\
It is clear that the algebra $(\oplus_{k = 1}^{\infty}
W^{(k)}(\{\omega^{(l)}\}_{l \leq k} ), \star)$ is associative and
unital.   Moreover, associativity also holds for contractions with
elements of the kind $(x)\star$ where $x \in
\mathcal{A}(x^{\alpha})^{tr}$. Elementary calculations suggest that
associativity is valid for arbitrary products in
$(\mathcal{A}(x^{\alpha})^{tr})(\{ \omega^{(k)} \}_{k \geq 0})^{tr}$
but a general proof seems rather involved\footnote{The following
identity is for sure useful
$$(a_1 \omega^{(s_1)}a_2 \ldots a_{n} \omega^{(s_n)} a_{n+1})\star
\omega^{(k)} =
 \omega^{(k')}(a_1 \omega^{(s'_1)}a_2 \ldots a_{n} \omega^{(s'_n)} a_{n+1})\star$$ where $k > 1$, $s'_p = s_p$ if $s_p < k + f^{k}(p)$ and $s_{p-1}$ otherwise.
  $f^{k}$ is defined inductively by $f^{k}(n) = 0$ and $f^{k}(p) = | \{ p < q \leq n | s_q < k + f^{k}(q)  \} |$ and finally, $k'= k + f^{k}(1)$ if $s_1 < k + f^{k}(1)$
  and $k + f^{k}(1) - 1$ otherwise.}. \\* \\
The property $\underbrace{(\omega^{(k)})\star \ldots
(\omega^{(k)})\star }_{k \textrm{ times}} w = w$ for any $w \in
W^{(k)}(\{\omega^{(l)}\}_{l \leq k} )$ suggests one to extend the
covariance rule using the generalized Jacobians $C^{(k)
\alpha}_{\quad \,\,\beta}(x';x)$, however this would not work.
First, one should worry about the covariance properties of the
differentials
$$\partial^{(k)}_{x^{\alpha}} : W^{(k-1)}(\{ \omega^{(l)} \}_{l \leq k-1}) \rightarrow W^{(k)}(\{ \omega^{(l)} \}_{l \leq
k})$$ or, when studying cohomology,
$$\partial^{(k)}_{x^{\alpha}} : Z^{k-1}(\{ \omega^{(l)} \}_{l \leq k-1}) \rightarrow Z^{k}(\{ \omega^{(l)} \}_{l \leq
k}).$$  To motivate the construction, let us write down the
following equalities:
$$\partial^{(2)}_{x^{\alpha}} \partial_{x^{\beta}} \Psi = (D(x';x)^{(2)
\kappa}_{\quad \alpha})\star \partial^{(2)}_{x'^{\kappa}}
(D(x';x)^{(1) \gamma}_{\quad \beta})\star \partial_{x'^{\gamma}}$$
$$ = \partial^{(2)}_{x^{\alpha}} (D(x';x)^{(1) \gamma}_{\quad \beta})\star
\partial_{x'^{\gamma}} + (D(x';x)^{(2) \kappa}_{\quad \alpha})\star
(D(x';x)^{(1) \gamma}_{\quad \beta})\star
\partial^{(2)}_{x'^{\kappa}} \partial_{x'^{\gamma}}$$ where the term
$\partial^{(2)}_{x^{\alpha}} (D(x';x)^{(1) \gamma}_{\quad \beta})$
is the usual gauge perturbation.  In order to have the right group
transformation properties, it is required that
$$D(x';x)^{(2) \kappa}_{\quad \alpha} \star D(x'';x')^{(2) \beta}_{\quad \kappa} = D(x'';x)^{(2) \beta}_{\quad
\alpha}$$ and
$$(D(x;x)^{(2) \kappa}_{\quad \alpha})\star \partial^{(2)}_{x^{\kappa}}
= \partial^{(2)}_{x^{\alpha}}$$ where in the first line equality
holds modulo derivatives of Cayley-Hamilton identities.  Moreover,
the equality
$$\left[ ( D(x'';x')^{(2) \gamma}_{\quad \kappa})\star ,
(D(x';x)^{(1) \alpha}_{\quad \beta})\star \right] = 0$$ needs to
hold on $W^{(2)}(\omega, \omega^{(2)})$. \\* \\  Clearly, a left
inverse $\rho^{(2)}$ of $\omega^{(2)}$ is needed, that is
$(\omega\rho^{(2)}) \star \omega^{(2)} = \omega$, $(\omega
\rho^{(2)}) \star \omega = \omega^{(2)} (\omega)\star$ and
$(\textrm{anything})\star$ commutes with any $\mathcal{A}$-valued
element.  This results in the definition
$$D(x';x)^{(2) \alpha}_{\quad \beta} =
\partial^{(2)}_{x^{\beta}}(x'^{\alpha})\rho^{(2)}$$ and from here, it is
easy to verify that all required properties hold: \begin{itemize}
\item $(x\omega^{(2)}y\rho^{(2)})\star \omega = \omega (x\omega
y)\star$ for any $\mathcal{A}$-valued expressions $x,y$,
\item The group property obviously holds since the chain rule for
the derivative $\partial^{(2)}_{x^{\alpha}}$ on $W^{(1)}(\omega)$ is
given by
$$\partial^{(2)}_{x^{\alpha}} = (\partial^{2}_{x^{\alpha}}
(x'^{\beta}) \rho^{(2)})\star \partial^{(2)}_{x'^{\beta}}$$ \item It
is obvious that two elements $(x\omega^{(2)}y\rho^{(2)})\star$ and
$(x'\omega y')\star$ commute on $W^{(2)}(\omega, \omega^{(2)})$
since the latter commutes with $\omega^{(2)}$.
\end{itemize}
Defining dual elements $d_{(2)}x^{\alpha}$ through
$$\partial^{(l)}_{x^{\alpha}} \star d_{(k)}x^{\beta} = \delta^{l}_{k}
\delta_{\alpha}^{\beta} \omega^{(k)} \rho^{(k)}$$ where $k,l = 1,2$
and $\rho^{(1)} = 1$ clearly is a coordinate invariant statement as
long as $$d_{(k)}x'^{\alpha} = d_{(k)}x^{\beta} \star
(\partial^{(k)}_{x^{\beta}}(x'^{\alpha}) \rho^{(k)}).$$  Therefore,
$(0,2)$ tensors are defined as $d_{(2)}x^{\alpha} \otimes
d_{(1)}x^{\beta} \star W_{\alpha \beta}$ where the coefficients
$W_{\alpha \beta} \in W^{(2)}(\omega, \omega^{(2)})$ and $(2,0)$
tensors are given by $V^{\alpha \beta} \star
\partial^{(1)}_{x^{\alpha}} \otimes \partial^{(2)}_{x^{\beta}}$
where $V^{\alpha \beta}$ is a linear combination of expressions of
the form $(a)\star (b\rho^{(2)})$ where $a,b \in
\mathcal{A}(x^{\alpha})^{tr}$. It is easy to verify that $V^{\alpha
\beta} \star W_{\alpha \beta} \in \mathcal{A}(x^{\gamma})^{tr}$ and
the transformation laws for $((a)\star (b\rho^{(2)}))^{\alpha
\beta}$ are given by $$((a)\star (b\rho^{(2)}))'^{\alpha \beta} =
((a)^{\gamma} \star
\partial_{x^{\gamma}} (x'^{\alpha})) \star ((b \rho^{(2)})^{\kappa}\star
(\partial^{(2)}_{x^{\kappa}}(x'^{\beta})\rho^{(2)}))$$ $(1,1)$
tensors are defined by putting the vector indices to the right; a
basis is given by
$$W_{\alpha}^{\quad \beta} = (W_{\alpha})\star(\omega a^{\beta}) $$ and the star
product of two $(1,1)$ tensors is again or the same type.
\\* \\
Generalization to higher derivatives leads to the introduction of
elements $\rho^{(k)}$ satisfying $(\omega \rho^{(k)})\star
\omega^{(k)} = \omega$, $(\omega \rho^{(k)})\star \omega^{(l)} =
\omega^{(l+1)}(\omega \rho^{(k-1)})\star$ for any $k>l$ and $(\omega
\rho^{(k)})\star \omega^{(l)} = \omega^{(l)}(\omega
\rho^{(k)})\star$ for $k < l$.  In general a basis of $(r,s)$
tensors is given by $$W_{\alpha_1 \ldots \alpha_r}^{\quad \beta_1
\ldots \beta_s} = W_{\alpha_1 \ldots \alpha_r} \star (\omega
a^{\beta_1}_1)\star (a^{\beta_2}_2 \rho^{(2)})\star \ldots
(a^{\beta_s}_s \rho^{(s)})$$ where $a^{\beta_k}_k \in
\mathcal{A}(x^{\alpha})^{tr}$ and notice that the order of the
$(a^{\beta_k}_k \rho^{(k)})$ is important.  It is easy to verify
that
$$(a_1)\star (a_2 \rho^{(2)})\star \ldots (a_s \rho^{(s)})\star
W_{t} \in W^{(t-s)}(\{\omega^{(l)}\}_{l \leq t-s})$$ for $t \geq s$.
 One needs to take care of contractions when the number of
contravariant indices of the left tensor $s$ is greater than the
number of covariant ones $t$ of the right tensor. In that case, one
obtains more complicated contravariant structures.  These rules
completely determine the tensor algebra; the next section deals with
derivation structures on it.
\subsection{Differential calculus.}  We generalize the
notion of Lie, exterior and covariant derivative; in the following paper integral calculus is developed which
suffices to define non-commutative
gravitational action principles. \\* \\
\underline{Lie derivative} \\
Let $x^{\alpha}(s)$ be a curve in $\mathcal{S}$ and denote by
$\frac{d}{ds}$ a derivation which acts upon $x^{\alpha}(s)$ as
$$\frac{dx^{\alpha}(s)}{ds}  = \frac{d}{ds}\left(tr(x^{\alpha}(s)
e_{\mu})\right)e^{\mu}$$ where the above expression clearly is
independent of the Pauli Cartan basis and by definition,
$\frac{dx^{\alpha}(s)}{ds}$ transforms as a vector. Let $\Psi$ be a
local diffeomorphism and note by $\Psi^{\alpha}$ the components of
$\Psi$ in a local chart around $\Psi(x^{\beta}(s))$.  Then, the
tangent vector to $\Psi^{\alpha}(x^{\beta}(s))$ is given by
$$\left(\Psi_{*}\frac{dx^{\gamma}(s)}{ds}\right)^{\alpha} := \frac{d\Psi^{\alpha}(x^{\beta}(s))}{ds} =
\frac{dx^{\gamma}(s)}{ds} \star
\partial_{x^{\gamma}}\Psi^{\alpha}(x^{\beta}(s))$$ and it is easy to
verify that the above expression is independent of the coordinate
system $x^{\alpha}$.  Let $V^{\alpha} \star
\partial_{x^{\alpha}}$ be a local vector field, then as usual it is possible to define integral curves of $$\frac{dx^{\alpha}(s,x^{\beta}_{0})}{ds} =
V^{\alpha}(x^{\beta}(s,x^{\gamma}_{0}))$$ and associated one
parameter group of local diffeomorphisms
$\Psi^{\alpha}_{t}(y^{\beta}) = x^{\alpha}(t,y^{\beta})$.  As usual
$$\mathcal{L}_{V}(W)(p) = \lim_{t \rightarrow 0} \frac{1}{t} \left(W(p) - \Psi_{t*} W(\Psi_{-t}(p)) \,
\right)$$ and \begin{eqnarray*} - \frac{d}{dt}\left(\Psi_{t*}
W(\Psi_{-t}(p))\right)^{\alpha} & = & - \frac{d}{dt} \,\,
W^{\beta}(x^{\gamma}(\Psi_{-t}(p)))\star
\partial_{x^{\beta}(\Psi_{-t}(p))}
\Psi^{\alpha}_{t}(x^{\gamma}(\Psi_{-t}(p))) \end{eqnarray*}
\begin{eqnarray*}
& = & V^{\kappa}(x^{\gamma}(\Psi_{-t}(p))) \star
\partial_{x^{\kappa}}(W^{\beta}(x^{\gamma}(\Psi_{-t}(p))))\star
\partial_{x^{\beta}(\Psi_{-t}(p))}
\Psi^{\alpha}_{t}(x^{\gamma}(\Psi_{-t}(p))) \, - \\
& & W^{\beta}(x^{\gamma}(\Psi_{-t}(p)))\star
\partial_{x^{\beta}(\Psi_{-t}(p))} \left( \frac{d}{dt}
\Psi^{\alpha}_{t} \right) (x^{\gamma}(\Psi_{-t}(p))) \, - \\ & &
W^{\beta}(x^{\gamma}(\Psi_{-t}(p)))\star \left(\frac{d}{dt}
x^{\kappa}(\Psi_{-t}(p)) \rho^{(2)} \right)\star
\partial^{(2)}_{x^{\kappa}(\Psi_{-t}(p))} \partial_{x^{\beta}(\Psi_{-t}(p))}
\Psi^{\alpha}_{t}(x^{\gamma}(\Psi_{-t}(p))).
\end{eqnarray*} Taking the limit $t \rightarrow 0$ results in $$(\mathcal{L}_{V}(W))^{\alpha}(p) =
V^{\beta}(x^{\gamma}(p))\star
\partial_{x^{\beta}}W^{\alpha}(x^{\gamma}(p)) -
W^{\beta}(x^{\gamma}(p)) \star \partial_{x^{\beta}}
V^{\alpha}(x^{\gamma}(p)) = \left[ V, W \right]^{\alpha}(p)$$ and it
is instructive to directly verify that this is indeed a vector.
 The Lie derivative of a scalar $\Psi$ is given by
$$\mathcal{L}_{V}(\Phi)(p) = - \frac{d}{dt}(\Phi(x^{\alpha}(\Psi_{-t}(p))))_{| t=0}=  V(\Phi)(p) \in \mathcal{A}(x^{\alpha})^{tr}$$
and given that $$\mathcal{L}_{V} T = - \frac{d}{dt}
\left(\Psi_{t*} T\right)_{| t =0}$$ where the push forward is
defined as in the commutative case, results in $$(\Psi_{t
*}Z)_{\alpha}(p) =
\partial_{x^{\alpha}(p)}\left( x^{\beta}(\Psi_{-t}(p)) \right)
\star Z_{\beta}\left( \Psi_{-t}(p) \right) $$ and therefore
$$(\mathcal{L}_{V}(Z))_{\alpha}(p) =
\partial_{x^{\alpha}(p)}(V^{\beta}(x^{\gamma}(p))) \star
Z_{\beta}(x^{\gamma}(p)) + (V^{\beta}(x^{\gamma}(p))\rho^{(2)})
\star
\partial^{(2)}_{x^{\beta}(p)}(Z_{\alpha}(x^{\gamma}(p)))$$ for any
one form $Z$ where covariance is understood to hold modulo
evaluation with $(a)\star$.  Generalizing to higher tensors results
in $$(\Psi_{t \star} W_{\alpha_1 \ldots \alpha_n})(p) = \left(
\partial^{(1)}_{x^{\alpha_1}(p)}x^{\beta_1}(\Psi_{-t}(p))
\right)\star \ldots \left(
\partial^{(n)}_{x^{\alpha_n}(p)}x^{\beta_n}(\Psi_{-t}(p))\rho^{(n)}
\right)\star W_{\beta_1 \ldots \beta_n}(\Psi_{-t}(p))$$ and
therefore $$\left( \mathcal{L}_{V} W \right)_{\alpha_1 \ldots
\alpha_n}(p) = \sum_{j = 1}^{n} \left(
\partial^{(j)}_{x^{\alpha_j}(p)}V^{\beta_j}(p)\rho^{(j)}
\right)\star W_{\alpha_1 \ldots \alpha_{j-1} \beta_j \alpha_{j+1}
\ldots \alpha_n}(p) + \left(V^{\beta}(p)\rho^{(n+1)}\right)\star
\partial^{(n+1)}_{x^{\beta}} W_{\alpha_1 \ldots \alpha_n}(p).$$
\underline{Exterior Calculus} \\ The usual external derivative $d$
serves to annihilate generalized gradients of functions and it is in
this vein that we define it.  Let $\sigma \in S_n$ and $(a_1)\star
(a_2\rho^{(2)})\star \ldots (a_n\rho^{(n)})$ be a $(n,0)$
contravariant tensor, then \begin{eqnarray*} (a_1)\star
(a_2\rho^{(2)})\star \ldots (a_n\rho^{(n)})\star P_{\sigma} & = &
(a_{\sigma(1)})\star \ldots (a_{\sigma(n)}\rho^{(n)})\star
\end{eqnarray*} Let $\Psi \in \mathcal{A}(x^{\alpha})^{tr}$, then
$d\Psi = d_{(1)}x^{\alpha} \star
\partial_{x^{\alpha}}\Psi$ and $$d^2 \Psi = d_{(1)}x^{\alpha_1} \otimes
d_{(2)}x^{\alpha_2} \star \left( \frac{1}{2!} \sum_{\sigma \in S_2}
\, sign(\sigma)P_{\sigma}
\partial^{(2)}_{x^{\alpha_{\sigma(2)}}}\partial^{(1)}_{x^{\alpha_{\sigma(1)}}}
\Psi \right) = 0$$ given the symmetry property of derivatives.
Therefore, given $W_{\alpha} \in W^{(1)}(\omega)$, $$d\left(
dx^{\alpha}\star W_{\alpha}\right) = d_{(1)}x^{\alpha_1} \otimes
d_{(2)}x^{\alpha_2} \star \left( \frac{1}{2!} \sum_{\sigma \in S_2}
\, sign(\sigma)P_{\sigma} \partial^{(2)}_{x^{\alpha_{\sigma(2)}}}
W_{\alpha_{\sigma(1)}} \right)$$ and it is a useful exercise to show
that the latter expression transforms as a covariant $(0,2)$ tensor.
\\* \\ \underline{Exercise} \\
$$
\frac{1}{2!} \sum_{\sigma \in S_2} \, sign(\sigma)P_{\sigma}
\partial^{(2)}_{x^{\alpha_{\sigma(2)}}} W_{\alpha_{\sigma(1)}} = $$
$$\frac{1}{2!} \sum_{\sigma \in S_2} \, sign(\sigma)P_{\sigma} \left(
\partial^{(2)}_{x^{\alpha_{\sigma(2)}}}\left(\partial_{x^{\alpha_{\sigma(1)}}}(x'^{\beta_1})\right)
\star W'_{\beta_1} +
\left(\partial_{x^{\alpha_{\sigma(1)}}}(x'^{\beta_1})\right)\star
\left(\partial^{(2)}_{x^{\alpha_{\sigma(2)}}}(x'^{\beta_2})\rho^{(2)}\right)\star
\partial^{(2)}_{x'^{\beta_2}} W'_{\beta_1}\right)$$ $$= \, \frac{1}{2!} \sum_{\sigma \in S_2} \, sign(\sigma)P_{\sigma} \left(
\left(\partial_{x^{\alpha_{\sigma(1)}}}(x'^{\beta_{\sigma(1)}})\right)\star
\left(\partial^{(2)}_{x^{\alpha_{\sigma(2)}}}(x'^{\beta_{\sigma(2)}})\rho^{(2)}\right)\star
\partial^{(2)}_{x'^{\beta_\sigma(2)}}
W'_{\beta_{\sigma(1)}}\right)$$ $$= \,
\left(\partial_{x^{\alpha_{1}}}(x'^{\beta_{1}})\right)\star
\left(\partial^{(2)}_{x^{\alpha_{2}}}(x'^{\beta_{2}})\rho^{(2)}\right)\star
\frac{1}{2!} \sum_{\sigma \in S_2} \, sign(\sigma)P_{\sigma}
\partial^{(2)}_{x'^{\beta_\sigma(2)}} W'_{\beta_{\sigma(1)}}
$$ \\ Obviously, one can extend elements of $W^{(k)}(\{ \omega^{(l)} \}_{l
\leq k})$, $k
> 1$, by adding the permutators $P_{\sigma}$ and we shall assume so
without further notice. In general,
$$d \left( d_{(1)}x^{\alpha_1}\otimes d_{(2)} x^{\alpha_2}\ldots \otimes d_{(k)}x^{\alpha_k} \star W_{\alpha_1 \ldots \alpha_k} \right)
= $$ $$ d_{(1)}x^{\alpha_1}\otimes d_{(2)} x^{\alpha_2}\ldots
\otimes d_{(k)}x^{\alpha_k} \otimes d_{(k+1)}x^{\alpha_{k+1}}\star
\left( \frac{1}{(k+1)!} \sum_{\sigma \in S_{k+1}} sign(\sigma)
P_\sigma
\partial^{(k+1)}_{x^{\alpha_{\sigma(k+1)}}} W_{\alpha_{\sigma(1)} \ldots
\alpha_{\sigma(k)}} \right)
$$ and an elementary calculation shows that $d^2 = 0$.  A $k$ form
field is defined by taking any $W_{\alpha_1 \ldots \alpha_k}$ and
anti-symmetrizing $$\frac{1}{k!} \sum_{\sigma \in S_k} sign(\sigma)
P_{\sigma} W_{\alpha_{\sigma(1)} \ldots \alpha_{\sigma(k)}}$$ and
our previous exercise teaches that this transforms again as a
$(0,k)$ tensor.  For any $(0,k)$ tensor $W_{\alpha_1 \ldots
\alpha_n}$, $\Psi_{\star} d W = d \Psi_{\star} W$ and therefore
$$\mathcal{L}_{V} d = d \mathcal{L}_{V}.$$
\\ \underline{Covariant derivative} \\
Remark first that the derivative of a tensor is simply defined by
deriving its components.  We present the covariant derivative in the
usual axiomatic way, define the associated connection, torsion and
curvature tensors and end with a discussion about the metric tensor.  Note from the outset that an infinity of
covariant differentials $\nabla^{(k)}_{\mathbf{V}_k}$ acting
respectively on $W^{(k-1)}(\{\omega^{(l)}\}_{l \leq k-1})$ where
$W^{0} = \mathcal{A}(x^{\alpha})^{tr}$ are needed.  For notational
simplicity we sometimes will use the symbol $\nabla_{\mathbf{V}}$
but it is understood that the correct grade ``k'' is used.  It is
useful to introduce the $\star$ algebra's $R^{(k)}$, $k>0$, spanned
by elements of the form $a\omega^{(k)}b\rho^{(k)}$ where $a,b \in
\mathcal{A}(x^{\alpha})^{tr}$ and $R^{(0)} =
\mathcal{A}(x^{\alpha})^{tr}$. For any point $p \in \mathcal{M}$ and
generalized vectorfield $\mathbf{V}_k = V^{\alpha}_k \star
\partial^{(k)}_{x^{\alpha}}$ of grade $k$ defined in a neighborhood of $p$ where $V^{\alpha}_k \in R^{(k)} \oplus (\mathcal{A}(x^{\beta})^{tr} \rho^{(k)})$,
we attach a differential operator $\nabla^{(k)}_{\mathbf{V}_k}$
satisfying
\begin{itemize}
\item $\nabla^{(k)}_{\mathbf{V}_k} \mathbf{W}$ is a tensor in the arguments,
that is for $f_k,g_k \in (\mathcal{A}(x^{\beta})^{tr}\rho^{(k)})$
and vector fields $\mathbf{X}_k, \mathbf{Y}_k$ with coefficients in
$R^{(k)}$, one has
$$\nabla^{(k)}_{(f_k \star \mathbf{X}_k + g_k \star \mathbf{Y}_k)} \mathbf{W} = f_k \star
\nabla^{(k)}_{\mathbf{X}_k}\mathbf{W} + g_k  \star \nabla^{(k)}_{
\mathbf{Y}_k}\mathbf{W}$$
\item $\nabla^{(k+1)}_{\mathbf{V}_{k+1}} (f_k \star \mathbf{X}_1 + g_{k} \star \mathbf{Y}_1 )  =
\mathbf{V}_{k+1}(f_k) \star \mathbf{X}_1 + \mathbf{V}_{k+1}(g_k)
\star \mathbf{Y}_1 + f_k \star \nabla^{(2)}_{\mathbf{V}_2}
\mathbf{X}_1 + g_k \star \nabla^{(2)}_{\mathbf{V}_2} \mathbf{Y}_1$
for $f_k,g_k \in R^{(k)}$, $k=0,1$ and $\mathbf{V}_2 = (V_{1}^{'
\alpha}\rho^{2}) \star
\partial^{(2)}_{x^{\alpha}}$ where $V_{1}^{'}$ is defined from $V_1$ by
 replacing $\omega$ by $\omega^{(2)}$ if necessary.
\end{itemize}
As usual, $\nabla_{\mathbf{V}}$ is a covariant derivative in the
direction of $\mathbf{V}$ at $p$. Defining the contraction
$\imath_{\mathbf{V}}$ of an $(r,s)$ tensor $W_{\alpha_1 \ldots
\alpha_s}^{\quad \beta_1 \ldots \beta_r}$ where $s > 0$ with
$V^{\alpha} \in \mathcal{A}(x^{\beta} )^{tr}$ as the $(r,s-1)$
tensor given by
$$(V^{\alpha})\star W_{\alpha \alpha_1 \ldots \alpha_{s-1}}^{\quad
\beta_1 \ldots \beta_r}$$ allows one to define the $(1,1)$ tensor
$$\nabla^{(1)} W =
 dx^{\alpha} \star W_{\quad ;\alpha}^{\beta} \star \partial_{x^{\beta}}$$
where $W^{\beta} \in W^{(0)}$ and therefore $W_{\quad
;\alpha}^{\beta} \in W^{(1)}(\omega)$.  Hence,
$$(\imath_{\mathbf{V}} \nabla^{(1)} W)^{\alpha} = V^{\beta} \star W_{\quad
; \beta}^{\alpha} = (\nabla^{(1)}_{\mathbf{V}} W)^{\alpha}.$$
Indeed, the first rule shows it suffices to consider the basis
vectors $\omega \star \partial_{x^{\alpha}}$ and by definition
$$\nabla^{(1)}_{(\omega \star \partial_{x^{\alpha}})} W =  W_{\quad ; \alpha}^{\beta} \star \partial_{x^{\beta}}$$ or $$ W_{\quad ;
\alpha}^{\beta} = \nabla^{(1)}_{(\omega \star
\partial_{x^{\alpha}})} W \star dx^{\beta}.$$  One defines a
connection $$\Gamma_{\alpha \beta}^{(1) \kappa} = \left(
\nabla^{(2)}_{((\omega^{(2)}\rho^{(2)}) \star
\partial^{(2)}_{x^{\alpha}})}( \omega \star \partial_{x^{\beta}} ) \right)
\star dx^{\kappa}$$ where the latter belongs to $W^{(2)}(\omega,
\omega^{(2)})$.  The following contraction rules are in place
$$\nabla^{(1)}_{\mathbf{V}} \mathbf{W} = \nabla^{(1)}_{V^{\alpha} \star \partial_{x^{\alpha}}} \left( W^{\beta} \star \partial_{x^{\beta}} \right) =
\mathbf{V}(W^{\beta}) \star \partial_{x^{\beta}}
 + (W^{\beta}) \star (V^{\alpha} \rho^{(2)})\star \nabla^{(2)}_{(\omega^{(2)}\rho^{(2)}) \star \partial^{(2)}_{x^{\alpha}}} \left( \omega \star \partial_{x^{\beta}}
 \right)$$ $$ = \mathbf{V}(W^{\beta}) \star \partial_{x^{\beta}}
 + (W^{\beta}) \star (V^{\alpha} \rho^{(2)})\star \Gamma_{\alpha \beta}^{(1)\kappa} \star \partial_{x^{\kappa}}$$
 $$ = \mathbf{V}(W^{\beta}) \star \partial_{x^{\beta}}
 + (V^{\alpha}) \star (W^{\beta})\star \Gamma_{\alpha \beta}^{(1)\kappa} \star
 \partial_{x^{\kappa}}$$ where the second and third line follow by different order of application of the calculational rules which proves its consistency.  Therefore
 $$W_{\quad ; \, \alpha}^{\beta} = \partial_{x^{\alpha}}(W^{\beta}) + W^{\kappa} \star \Gamma_{\alpha
 \kappa}^{(1)\beta}$$ and the transformation rules are determined by
\begin{eqnarray*} \Gamma_{\alpha \beta}^{(1) ' \kappa} & = & \left( \nabla^{(2)}_{((\omega^{(2)} \rho^{(2)}) \star
\partial^{(2)}_{x'^{\alpha}})}( \omega \star \partial_{x'^{\beta}} ) \right)
\star dx'^{\kappa} \\
& = & \left( \nabla^{(2)}_{
(\partial^{(2)}_{x'^{\alpha}}(x^{\delta}) \rho^{(2)}) \star
(\omega^{(2)} \rho^{(2)}) \star
\partial^{(2)}_{x^{\delta}}} (\partial_{x'^{\beta}} (x^{\gamma}) \star
\omega \star \partial_{x^{\gamma}} ) \right) \star dx^{\lambda} \star \partial_{x^{\lambda}}(x'^{\kappa}) \\
& = &
\partial^{(2)}_{x'^{\alpha}} \partial_{x'^{\beta}}(x^{\lambda}) \star \partial_{x^{\lambda}}(x'^{\kappa})
+ (\partial_{x'^{\beta}} (x^{\gamma}))\star
(\partial^{(2)}_{x'^{\alpha}}(x^{\delta}) \rho^{(2)}) \star
\Gamma_{\delta \gamma}^{(1) \lambda} \star
\partial_{x^{\lambda}}(x'^{\kappa})
\end{eqnarray*}
By definition $W_{\quad ; \, \alpha}^{\beta}$ transforms as a
$(1,1)$ tensor; however it instructive to verify this explicitly. As
usual, the difference of two connections transforms as a tensor.
Prior to extending the covariant derivative to general tensors, we
apply this differential calculus to the algebra $R^{n}$. \\* \\
\underline{Preference of flat space time.} \\
It is obvious that the covariant calculus developed above does
\underline{not} coincide with the standard calculus on $R^{n}$ which
is now to be seen as a one dimensional $(R^n,R)$ manifold.  As
before, $\partial_{x_j} = (0, \ldots ,0,1_j,0,\ldots,0) \star
\partial_x$ and
$$\Gamma_{jk}^{l} = (0, \ldots ,0,1_k,0,\ldots,0)\star ((0, \ldots
,0,1_j,0,\ldots,0)\rho^{(2)})\star \Gamma_{xx}^{(1)x}\star (0,
\ldots ,0,1_l,0,\ldots,0)$$ is only nonzero in case $j=k=l$.  This
is so because there is only one direction in $R^n$ while there are
$n$ in the standard view.  Therefore, only \emph{flat} abelian
space-times exist since a coordinate transformation
$$\frac{\partial^{2} x'}{\partial x \partial x} = -
\Gamma_{xx}^{(1) x}$$ is sufficient to make the connection vanish
everywhere.  Therefore, small deviations from flatness might be seen as
equivalent to small deviations from commutativity which is again
equivalent to the existence of matter. \\* \\
We now generalize the covariant derivative to general tensors of
type $(r,s)$:
\begin{itemize}
\item If $T$ is a tensor field of type $(r,s)$, then $\nabla T$ is
of type $(r,s+1)$,
\item $\nabla$ is linear and commutes with contractions,
\item $\nabla$ shifts through tensor products
\item $\nabla f = df$ for any $f \in \mathcal{A}(x^{\alpha})^{tr}$.
\end{itemize}
Specifically, for a one form $dx^{\alpha} \star (\omega)$, this
means that $$0 = \nabla^{(2)}_{(\omega^{(2)}\rho^{(2)})\star
\partial^{(2)}_{x^{\alpha}}} \left((\omega) \star
\partial^{(1)}_{x^{\beta}} \star d_{(1)} x^{\kappa} \star (\omega) \right) = \Gamma_{\alpha \beta}^{(1) \kappa} +  (\omega) \star
\partial^{(1)}_{x^{\beta}} \star \nabla^{(2)}_{(\omega^{(2)}\rho^{(2)})\star
\partial^{(2)}_{x^{\alpha}}}(d_{(1)} x^{\kappa} \star \omega)$$ and
therefore $$\nabla^{(2)}_{(\omega^{(2)}\rho^{(2)})\star
\partial^{(2)}_{x^{\alpha}}}(d_{(1)} x^{\kappa} \star \omega) = - d_{(1)} x^{\gamma} \star \Gamma_{\alpha
\gamma}^{(1) \kappa}$$ implying $$\nabla^{(2)}_{\mathbf{V}_2}
\mathbf{W} = dx^{\gamma} \star \left( (V_{2}^{\alpha}) \star
\partial^{(2)}_{x^{\alpha}}(W_{\gamma}) - (V_{2}^{\alpha}) \star \Gamma_{\alpha
\gamma}^{(1) \kappa} \star W_{\kappa} \right).$$  For higher tensors
we have to take care of the $\omega^{(k)}$'s and $\rho^{(k)}$'s
being in the right place.  Indeed, an identical calculation as
before reveals that $$\nabla^{(k)}_{(\omega^{(k)}\rho^{(k)})\star
\partial^{(k)}_{x^{\alpha}} } (dx_{(k-1)}^{\beta} \star (\omega^{(k-1)}\rho^{(k-1)})) = -
d_{(k-1)}x^{\kappa} \star \Gamma_{\alpha \kappa}^{(k-1) \beta} $$
where we \underline{demand} now that $\Gamma_{\alpha \kappa}^{(k-1)
\beta}$ equals $\Gamma_{\alpha \kappa}^{'(1) \beta}\rho^{(k-1)}$
where $\Gamma_{\alpha \kappa}^{'(1) \beta}$ is constructed from
$\Gamma_{\alpha \kappa}^{(1) \beta}$ by replacing $\omega$ with
$\omega^{(k-1)}$ and $\omega^{(2)}$ with $\omega^{(k)}$.  Therefore,
for any $(0,s)$ tensor we have $$(\nabla^{(s+1)}_{\mathbf{V}_{s+1}}
W)_{\alpha_1 \ldots \alpha_s} = \mathbf{V}_{s+1}(W_{\alpha_1 \ldots
\alpha_s}) \, - \, \sum_{j=1}^{s} (V_{j+1}^{\beta})\star
\Gamma_{\beta \alpha_j}^{(j) \gamma} \star W_{\alpha_1 \ldots
\alpha_{j-1} \gamma \alpha_{j+1} \ldots \alpha_{s}}$$ where
$V_{j+1}^{\beta} \in \mathcal{A}(x^{\alpha})^{tr} \rho^{(j+1)}$.   Covariance is
readily verified since elements of different $R^{(k)}$'s $k \leq s$
commute in front of a $W_{\alpha_1 \ldots \alpha_s}$.  We shall
further generalize (a) the geodesic equation (b) the notions of
torsion and curvature. \\* \\
\underline{Geodesics, torsion and curvature} \\
A geodesic is a line in $\mathcal{S}$ such that the tangent vector
is parallel transported along itself.  That is, given $V^{\alpha} =
\frac{d x^{\alpha}(s)}{ds}$, one defines
$$V^{\alpha} \star V^{\beta}_{\quad ; \, \alpha} = V^{\alpha} \star \partial_{\alpha} V^{\beta} + V^{\alpha }\star V^{\kappa} \star \Gamma_{\alpha \kappa}^{\beta} = 0$$ and $s$ is an affine parametrization,
determined upon a transformation $as + b$.  The construction of the
exponential map and normal coordinates proceeds as usual and may be
left as an exercise to the reader.  Given two vector fields
$\mathbf{X}, \mathbf{Y}$, the torsion tensor is defined as
$$\mathbf{T}(\mathbf{X},\mathbf{Y}) = \nabla_{\mathbf{X}} \mathbf{Y}
- \nabla_{\mathbf{Y}} \mathbf{X} - \left[ \mathbf{X} , \mathbf{Y}
\right]$$ and we explicitly prove it is a tensor:
\begin{eqnarray*}
\mathbf{T}((a)\star \mathbf{X}_1, (c)\star \mathbf{Y}_1) & = &
\nabla^{(1)}_{(a)\star \mathbf{X}_1}((c)\star \mathbf{Y}_1) -
\nabla^{(1)}_{(c)\star \mathbf{Y}_1}((a)\star \mathbf{X}_1) - \left[
(a)\star \mathbf{X}_1, (c)\star \mathbf{Y}_1 \right] \\ & = &
(a)\star (c)\star \nabla^{(2)}_{\mathbf{X}_2} \mathbf{Y}_1 +
(a)\star \mathbf{X}_1(c) \star \mathbf{Y}_1 - (c)\star (a)\star
\nabla^{(2)}_{\mathbf{Y}_2} \mathbf{X}_1 - \\ & & (c)\star
\mathbf{Y}_1(a) \star \mathbf{X}_1 - (a)\star (c) \star \mathbf{X}_2
\mathbf{Y}_1 + (c)\star (a)\star \mathbf{Y}_2 \mathbf{X}_1
\\ & & - (a)\star \mathbf{X}_1(c)\star \mathbf{Y}_1 + (c) \star
\mathbf{Y}_1(a) \star \mathbf{X}_1
\end{eqnarray*} and the last line equals  $$(a)\star (c)\star \nabla^{(2)}_{\mathbf{X}_2}
\mathbf{Y}_1 - (c)\star (a)\star \nabla^{(2)}_{\mathbf{Y}_2}
\mathbf{X}_1 - (a)\star (c) \star \mathbf{X}_2 \mathbf{Y}_1+
(c)\star (a)\star \mathbf{Y}_2 \mathbf{X}_1
$$ which is what we needed to prove.  One notices that due to the order in which the vector
fields are
 applied, the coefficients shift through in different ways.  Therefore, in a component
notation, it is useful to introduce the following $$T_{\beta_1 \beta_2} = \sum_{\sigma \in S_2} P_{\sigma} T^{\sigma}_{\beta_1 \beta_2}$$ where the $T^{\sigma}_{\alpha_1 \alpha_2}$ are not tensors themselves.  The
torsion tensor vanishes if and only if
$$(a)\star (b\rho^{(2)})\star
\nabla^{(2)}_{(\omega^{(2)}\rho^{(2)})\star
\partial^{(2)}_{x^{\alpha}}} \left ((\omega) \star
\partial_{x^{\beta}} \right)\, - \, (b)\star (a \rho^{(2)}) \star \nabla^{(2)}_{(\omega^{(2)}\rho^{(2)})\star
\partial^{(2)}_{x^{\beta}}} \left ((\omega) \star
\partial_{x^{\alpha}} \right) \, = \, 0$$
or $$(b)\star (a) \star \Gamma_{\alpha \beta}^{(1) \kappa} = (a)
\star (b)\star \Gamma_{\beta \alpha}^{(1) \kappa}$$ implying that
the geodesic equation captures the entire connection.  The usual
relationship between torsion free connections and Lie derivatives
survives, indeed
$$ \mathcal{L}_{\mathbf{V}_0} (\mathbf{W}_0) = \left[ \mathbf{V}_0 ,
\mathbf{W}_0 \right] = \nabla^{(1)}_{\mathbf{V}_0} \mathbf{W}_{0} -
\nabla^{(1)}_{\mathbf{W}_{0}} \mathbf{V}_{0} $$ and therefore
$$ \left( \mathcal{L}_{\mathbf{V}_0} (\mathbf{W}_0) \right)^{\alpha} = V^{\beta} \star W^{\alpha}_{\quad ; \, \beta} - W^{\beta} \star V^{\alpha}_{\quad ; \, \beta}.$$
The usual relationship between exterior calculus and torsionfree
covariant derivatives is also valid due to the presence of the
permutators $P_{\sigma}$.   We proceed now with the definition of
the curvature tensor $\mathbf{R}(\mathbf{X}, \mathbf{Y})
\mathbf{Z}$.  The latter coincides with the usual expression
$$\mathbf{R}(\mathbf{X}, \mathbf{Y})
\mathbf{Z} = \nabla_{\mathbf{X}} \left( \nabla_{\mathbf{Y}}
\mathbf{Z} \right) - \nabla_{\mathbf{Y}} \left( \nabla_{\mathbf{X}}
\mathbf{Z} \right) - \nabla_{\left[ \mathbf{X}, \mathbf{Y} \right]}
\mathbf{Z}$$ and one can verify in the same way that $\mathbf{R}$ is
a $(1,3)$ tensorfield.  In the same vein as before $$\mathbf{R}((a)\star \mathbf{X}_1,(b)\star \mathbf{Y}_1) \mathbf{Z}= (a)\star (b)\star \nabla^{(2)}_{\mathbf{X}_2} \nabla_{\mathbf{Y}_1} \mathbf{Z} \, - (b)\star (a)\star \nabla^{(2)}_{\mathbf{Y}_2} \nabla_{\mathbf{X}_1} \mathbf{Z}\, - \nabla_{(a)\star(b)\star \mathbf{X}_2 \mathbf{Y}_1 \, - (b)\star(a)\star \mathbf{Y}_2 \mathbf{X}_1 }\mathbf{Z}.$$
In a component basis, one obtains that
$$(\mathbf{R}(\mathbf{X}, \mathbf{Y})\mathbf{Z})^{\alpha} = (Z^{\kappa}) \star (Y^{\gamma}\rho^{(2)}) \star (X^{\beta} \rho^{(3)}) \star \left( \nabla^{(3)}_{(\omega^{(3)}\rho^{(3)}) \star \partial^{(3)}_{x^{\beta}}} \left( \nabla^{(2)}_{(\omega^{(2)}\rho^{(2)}) \star \partial^{(2)}_{x^{\gamma}}} (\omega) \star \partial_{x^{\kappa}} \right) \right)^{\alpha} \,- $$ $$(Z^{\kappa}) \star (X^{\beta}\rho^{(2)}) \star (Y^{\gamma} \rho^{(3)}) \star \left( \nabla^{(3)}_{(\omega^{(3)}\rho^{(3)}) \star \partial^{(3)}_{x^{\gamma}}} \left( \nabla^{(2)}_{(\omega^{(2)}\rho^{(2)}) \star \partial^{(2)}_{x^{\beta}}} (\omega) \star \partial_{x^{\kappa}} \right) \right)^{\alpha} $$
or, calculated directly
\begin{eqnarray*}
(\mathbf{R}(\mathbf{X}, \mathbf{Y})\mathbf{Z})^{\alpha} & = & (Y^{\beta})\star (X^{\kappa}\rho^{(2)}) \star Z^{\alpha}_{; \, \beta \kappa} - (X^{\kappa})\star (Y^{\beta}\rho^{(2)}) \star Z^{\alpha}_{; \, \kappa \beta}.
\end{eqnarray*}
Considering the previous formula, one obtains that $$\left( \mathbf{R}((\omega)\star \partial_{x^{\beta}}, (\omega)\star \partial_{x^{\gamma}} ) (\omega)\star \partial_{x^{\kappa}} \right)^{\alpha} = \partial^{(3)}_{x^{\beta}} \left( \Gamma^{(1) \alpha}_{\gamma \kappa} \right) \, - \, P_{(23)} \partial^{(3)}_{x^{\gamma}} \left( \Gamma^{(1) \alpha}_{\beta \kappa} \right) \, + $$ $$ \, \Gamma^{(1) \lambda}_{\gamma \kappa} \star \Gamma^{(1) \alpha}_{\beta \lambda} \, - \, P_{(23)} \Gamma^{(1) \lambda}_{\beta \kappa} \star \Gamma^{(1) \alpha}_{\gamma \lambda} = R^{\alpha}_{\, \kappa \beta \gamma}.$$
Clearly, the curvature tensor is antisymmetric in the arguments $\mathbf{X}$ and $\mathbf{Y}$:
$$\mathbf{R}(\mathbf{X}, \mathbf{Y})\mathbf{Z} = - \mathbf{R}(\mathbf{Y}, \mathbf{X})\mathbf{Z}$$ and moreover, the first Bianchi identity holds $$\mathbf{R}(\mathbf{X}, \mathbf{Y})\mathbf{Z} + \mathbf{R}(\mathbf{Z}, \mathbf{X})\mathbf{Y} + \mathbf{R}(\mathbf{Y}, \mathbf{Z})\mathbf{X} = 0$$ due to the connection being torsion free and the Jacobi identity and in component form: $$R^{\alpha}_{\, \kappa \beta \gamma} + P_{\left(
                                             \begin{array}{ccc}
                                               1 & 2 & 3 \\
                                               2 & 3 & 1 \\
                                             \end{array}
                                           \right)
}R^{\alpha}_{\, \gamma \kappa \beta} + P_{\left(
                                            \begin{array}{ccc}
                                              1 & 2 & 3 \\
                                              3 & 1 & 2 \\
                                            \end{array}
                                          \right)
}R^{\alpha}_{\, \beta \gamma \kappa}$$  The second Bianchi identity $$\nabla_{\mathbf{W}} \mathbf{R}(\mathbf{X}, \mathbf{Y})\mathbf{Z} + \nabla_{\mathbf{Y}} \mathbf{R}(\mathbf{W}, \mathbf{X})\mathbf{Z} + \nabla_{\mathbf{X}} \mathbf{R}(\mathbf{Y}, \mathbf{W})\mathbf{Z} $$ likewise holds and it is left as an easy exercise to write it in component form.
 \\* \\ \underline{Metrics} \\ Obviously, we are interested only in generalized hyperbolic
structures since we believe physics to be causal.  The interesting part is that the
noncommutative notion of local causality can violate Bell's theorem and indeed, we shall
derive quantum mechanical predictions later on.  Before proceeding however, it is necessary
to tell something about bases in $T^{\star} \mathcal{M}$ : $$ E^{\alpha}_{j} \star \partial_{x^{\alpha}}$$
$j,\alpha : 1 \ldots M$ and $E^{\alpha}_{j} \in W^{(1)}(\omega)$ is a basis if and only if for any vector $\mathbf{W}_{0}$ there exist $a_j \in \mathcal{A}$ such that $$\mathbf{W}_{0} = \sum_{j=1}^{M} a_j \star \mathbf{E}_j.$$  As before, it is easy to prove that there exists a co-basis $\mathbf{E}^{j}$ such that $$\mathbf{E}_j \star \mathbf{E}^k = \delta_{j}^{k} \omega$$ $$ E_{\alpha}^j \star E_{j}^{\beta} = \delta^{\beta}_{\alpha}\omega .$$  Indeed it is sufficient to note that $\mathbf{E}_j$ is a basis if and only if the $(NM)\times(NM)$ vielbein $$e^{\mu \alpha}_{\nu j} = tr(e_{\nu} \star E^{\alpha}_j e^{\mu})$$ is invertible.  Obviously, only $NM^2$ of the complex invertible $(NM)\times(NM)$ matrices can be written in this way and $$(\omega)\star \partial_{x^{\alpha}} = e_{\kappa} \left( e^{-1} \right)^{\kappa j}_{\gamma \alpha} tr(\omega e^{\gamma}) \star \mathbf{E}_j.$$
Taking $(1)\star \mathbf{E}_j \equiv \widetilde{\mathbf{E}}_j$ provides us with a vielbein satisfying $$\widetilde{\mathbf{E}}_j \star \mathbf{E}^k = \delta_{j}^{k}.$$  Obviously, given $\widetilde{\mathbf{E}}_j$, there is no unique inverse vielbein; the residual freedom is given by matrices $A^{\kappa k}_{\mu \alpha}$ satisfying
 $$A^{\kappa k}_{\mu \alpha} tr((1)\star \mathbf{E}^{\alpha}_j e^{\mu})= 0$$ for all $\kappa, k, j$.  For example, put $\mathbf{E}_j = (\omega a) \star \partial_{x^{j}}$ where $a$ is invertible; consider elements $b_k, c_k$ such that $b_k c_k \neq 0 = b_k a c_k$, then $$ \mathbf{E}^{k} = dx^{k} \star (b_k \omega c_k)$$ kills off all $\widetilde{\mathbf{E}}_j$.  Given a set of basis vectors, it is possible to construct a Lorentzian tensor as $$g_{\widetilde{\alpha} \beta} = \eta_{jk} E_{\widetilde{\alpha}}^{\star \, (2) \, j}E_{ \beta}^{k}.$$  Indeed, it is easy to verify that $$
 (E^{\beta}_j)\star (E^{\star \, \widetilde{\alpha}}_{(2) \, k}\rho^{(2)})\star g_{\widetilde{\alpha} \beta} = \eta_{jk} \omega^{(2)}\omega$$ and $$ \eta^{jk} (E^{\star \, \widetilde{\alpha}}_{(2) \, j})\star g_{\widetilde{\alpha} \beta}\star E^{\lambda}_k = \delta_{\beta}^{\lambda} \omega^{(2)}\omega.$$  The last line points out there is no canonical way to write the inverse metric as $g^{\alpha \widetilde{\beta}}$ and taking the inverse operation shall occur as above.  It is important to note that we have formally doubled the number of variables $x^{\alpha}$, if necessary, by adding $x^{\star \widetilde{\alpha}}$ just like this happens in complex geometry.  Obviously, we only allow for analytic diffeomorphisms, that is those which do not mix up the $x^{\alpha}$ and $x^{\star \widetilde{\alpha}}$. \\* \\ \underline{Remark}: In general, one would expect a Lorentz metric $g$ to be defined as a non degenerate hermitian $(0,2)$ tensor with signature $(- + + \ldots +)$; that is for any vector fields $\mathbf{X}, \mathbf{Y}$ one has that $$g(\mathbf{X}, \mathbf{Y})^{\star} = g(\mathbf{Y}^{\star}, \mathbf{X}^{\star})$$ and therefore $g(\mathbf{X}^{\star}, \mathbf{X})$ is a hermitian operator.  Non degeneracy of $g$ can be defined as follows: for any basis $\mathbf{E}_j$ one has that $g(\mathbf{E}^{\star}_j, . )$ is a basis in $T_{\star} \mathcal{M}$.  However, in order to define signature $(- + + \ldots +)$, one requires an analog of the Gram Schmidt or spectral theorem.  Given that we shall directly work with the vielbein, we postpone this issue for later.  \\* \\ This is a suitable point to stop the development of non commutative calculus since what we have now suffices to derive a bunch of interesting physics results.  The follow-up paper will treat Levi-Civita connections, integral calculus and an equivalent of Stokes theorem; that is, all the tools needed for defining an action principle.
\section{Non commutative Physics}
\subsection{Selection of kinematical setting.}
The aim of this section is to formulate a few powerful mathematical
and physical principles determining uniquely the kinematical
setting.  Let us start with a few philosophical notes; a material
entity is an active operation on the kinematical
structure. Usually, one considers tensor fields as dynamical
entities, however the latter belong to the passive
kinematical setting. Therefore, an action principle should be
operator valued, that is it contains successive actions
of covariant derivatives $\nabla_{\mathbf{V}}$ or multiplication operators.  A key lesson from the previous chapter is that in order to do geometry, (a) one needs needs to specify the algebra $\mathcal{A}$ (b) a specific sub-algebra $\mathcal{B}$ and associated generators $x^{\alpha}$ so that $\mathcal{B}(1,x^{\alpha}) = \mathcal{A}$.  Now, we can  even go further and introduce real structures $r_j$ associated to $\mathcal{A}$ resulting in fundamental coordinates $$x^{\alpha}, {x^{\alpha}}^{r_j}.$$  The idea is then that the fundamental degrees of freedom correspond to real structures of a \emph{super structure} with as local diffeomorphism group $\textrm{Diff}(\mathcal{S})$ where $\mathcal{S}$ is the space of generating $M$-tuples of $\mathcal{A}$ over $\mathcal{B}$.  A local (over $\mathcal{S}$) gauge degree of freedom consists in rotating the real structures themselves preserving some desirable algebraic features, this leads to natural bundles and as an example we derive $U(1) \otimes SU(3)$.  More in particular: \begin{itemize}
\item  (I) The fundamental physical entities correspond to nilpotent
operators implying that real
coordinates must take value in the generating set of a maximal real unital Grassman algebra $\mathcal{N}$.  In order to speak about the bosonic and fermionic part of a coordinate, the latter must be endowed with a reversion $r$ as well as an involution $\star$.  That is, $\mathcal{N}$ can be decomposed as $$\mathcal{N} = E \oplus O$$ where $E, O$ are self adjoint subalgebra's of even and odd $r$ parity respectively.  $E$ is the center of $\mathcal{N}$ and elements in $O$ are mutually anti commuting.
\item (II) $\mathcal{N}$ is extended over $H \oplus H$, where $H$ denotes the complex quaternions; the reason for this is that $H$ carries the Lorentz group, and the
 latter has a natural left and right representation on $H$ itself\footnote{See the notion of Vector Algebra or, Vecalg explained below.}.
\item (III) The super coordinates $x^{\alpha}$ generate $\mathcal{A} = \mathcal{B} \otimes \mathcal{N}$ over $\mathcal{B}$.  We furthermore demand that the traces of generators constitute four independent real coordinates.
\item (IV) Fundamental coordinates are constructed from the $x^{\alpha}$ by applying \emph{any} maximal set of involutions respecting the complex algebraic structure in $\mathcal{B}$.
\item (V) The fundamental fields determine a unique Lorentzian
causal structure.
\end{itemize}
(I) implies that fundamental fields are fermions ; indeed the Pauli
principle holds (at a linearized level) : consider
$N_1$ and $N_2$, then the sum is
(approximately) a Fermi field if and only if $$\{ N_1 , N_2 \} \approx 0.$$  (I) to (III)
imply that $\mathcal{B}$ is the algebra $H \oplus H$ and $\mathcal{N}= GRAS_{(4)}(R)$ where the latter is formed by unity, four anticommuting elements $a_{\alpha}$ and their
conjugates $a_{\alpha}^{*}$ satisfying $$\{a_{\alpha} ,
a_{\beta}^{*} \} = 0.$$ $\mathcal{N}$ is $256$ dimensional over $R$
as a vector space.  $\left(H \oplus H \right) \otimes GRAS_{(4)}(R)$ is generated by eight
elements $x^{\alpha}, x^{\alpha \, \dag}$ given that the multiplying algebra is considered to be $H \oplus H$.  \\* \\
\underline{Remark} \\ Note that we have only fixed the local structure; therefore, it might very well be that the non-abelian directions of global space-time $\mathcal{M}$ are compactified in order to obtain the desired stability properties, as elucidated in the introduction.  We have nothing to say on this matter prior to having developed and studied stability properties of local dynamics.  \\* \\  $\mathcal{A}$ allows for six independent involutions: (i) complex conjugation $\overline{x}$, (ii) the canonical involution $x^{\star}$ and reversion $x^{r}$ on $GRAS_{(4)}(R)$ and (iii) the three involutions $\overline{x}^k$ corresponding to $q_k \rightarrow - q_k$.
Given the usual quaternion base $q_{\kappa}$, we may write that $x^{\alpha} = q^L_{\kappa} x^{\alpha \, \kappa}_L + q^R_{\lambda} x^{\alpha \, \lambda}_R  $
where the $x^{\alpha \, \kappa}_L, x^{\alpha \, \kappa}_R$ are $C \otimes \mathcal{N}$ valued.
Axiom (IV) implies that real coordinates are constructed by applying the complex conjugation,
reversion and Grassman conjugation $\star$ to the $x^{\alpha \, \kappa}_L, x^{\alpha \, \kappa}_R$.
However, we find it more suitable to work with the set $$x^{\alpha \, \pm \, \kappa}_{H}, \overline{x^{\alpha \, \pm \, \kappa}}_{H},
 {x^{\alpha \, \pm}_{\kappa \, H}}^{\dag}, {\overline{x^{\alpha \, \pm }_{\kappa \, H}}}^{\dag}$$ where the
 index $\pm$ indicates even or odd parity under $r$ and $\dag$ is the composition of
 $\star$, the ordinary complex and quaternion conjugation and $H$ takes values in $\{ L,R \}$.  Since the coordinates $x^{\alpha}$ take
value in the generating set of the algebra $\left( H \oplus H \right) \otimes \mathcal{N}$, there exist only four real central coordinates.
For any $\alpha$, one counts exactly $16$ Fermi coordinates implying
$16$ left handed and $16$ right handed complex spinors which is exactly the number of left and right handed
particles and anti-particles in one standard model generation.  Axiom (V) implies there exist four fundamental fields
$E^{j}$ where $j=1 \ldots 4$ as well as their conjugate
fields which cook up a Lorentz metric as follows
$$\eta_{jk} E^{\dag (2)j} E^{k}$$ implying there are $3$ ``space-like generations'' and one temporal
generation.  The $E^{j}$, $j = 1 \ldots 3$, are conjectured to constitute the generations of the standard
model and $E^{0}$ gives an entirely different perspective upon
the notion of time.  We finish this section by making some comments
about the measurement problem in quantum mechanics which should be solved by any generalized local realist attempt.
  Obviously,
 waves do not interact since they constitute solutions of the linearized
equation and therefore only superpose.
Hence, measurement of a particle is due to a strong non linear
effect requiring the presence of a vortex. Indeed, an
electromagnetic wave can only interact with an electron in an atomic
orbit if there is a light vortex to bound and resonate with.  An
electron black hole corresponds with a singularity at the level of
fundamental fields, while a photon vortex corresponds to a
singularity at the level of gauge fields.  The bound
state statistics is exactly given by the Fermi property formulated
above since particles inherit the quantum numbers of waves.  Let me stress that this
isn't in \emph{any} sense in contradiction with the superposition principle of quantum theory
just like the nonlinear trajectories in Bohmian mechanics and the corresponding solution of the
cat problem aren't. \\* \\
The reader should appreciate the magic of the number four.
The latter equals : (a) the maximal dimension of a real division
algebra (b) the minimal dimension which allows for local
gravitational degrees of freedom (c) the number of real, central
components of coordinates in nature. \subsection{Special geometrical setting preserving algebraic properties.}
Now, we turn to the algebra at hand and develop (generalized) quaternionic super-space\footnote{Our calculus isn't quite superspace calculus, merely by the choice of bosonic variables.}
 calculus.  That is,
we abandon the Leibnitz rule and retrieve Taylor's rule in a different way.
Given that the latter holds in both
cases, there is an obvious embedding of the algebra-specific calculus into our more general
framework.  Let $\overline{x}$ and $\widetilde{x}$ be the complex and quaternion conjugate of $x$ respectively and
note by $q_{\kappa}$ the complex quaternions; that is $q_{\kappa} \equiv (1, i q_{k})$ where
$q_{k} q_{l} = \epsilon_{klm} q^{m \, \dag} - \delta_{kl}$.  Given that for any $a$, the following holds
$$(a)\star \partial_{x^{\alpha}} \overline{x^{\beta}} = \delta_{\alpha}^{\beta} \overline{a}$$ one deduces that
$$\frac{1}{2} \left( (a)\star - i (ia)\star \right)\partial_{x^{\alpha}} x^{\beta} = \delta_{\alpha}^{\beta} a$$
and
$$\frac{1}{2} \left( (a)\star - i (ia)\star \right)\partial_{x^{\alpha}} \overline{x^{\beta}} = 0.$$
Likewise, the partial derivatives of the bar coordinates in the direction of $a$ are given by
 $\frac{1}{2} \left( (a)\star + i (ia)\star \right)\partial_{x^{\alpha}}$.
Define the operators
$\widehat{\partial}_{x^{\alpha \, \pm \, \kappa}}$ by the following equation
$$a^{\pm}\widehat{\partial}_{x^{\alpha \, \pm \, \kappa}} =
 \frac{1}{2} \left( (a^{\pm}q_{\kappa})\star \,
- \, i (ia^{\pm}q_{\kappa})\star \right)\partial_{x^{\alpha}}$$ and likewise for the complex
 and $\dag$ conjugates; $a^{\pm}$ belongs to the subalgebra's $C \otimes E$ and $C \otimes O$
 respectively.  Given that $$x^{\alpha \, \pm \, k} = \frac{1}{8} \left( \{ x^{\alpha}
 - \widetilde{x^{\alpha}} , iq^{k \, \dag} \}  \,
\pm \, \{x^{\alpha \, r}
 - \widetilde{x^{\alpha \, r}} , iq^{k \, \dag}  \}  \right)$$ and $$x^{\alpha \, \pm \, 0} =
\frac{1}{4} \left( (x^{\alpha}
 + \widetilde{x^{\alpha}}) \,
\pm \, (x^{\alpha \, r}
 + \widetilde{x^{\alpha \, r}}) \right)$$ it is easy to prove that the $$\widehat{\partial}_{x^{\alpha \, - \, \kappa}}$$
are the usual anti-commuting derivatives, with a residual ambiguity, and likewise so for
the complex and $\dag$ conjugate.  The ambiguity resides in the use of the bosonic element $$a_1 \ldots a_4 a^{\star}_1 \ldots a^{\star}_4$$
which vanishes upon multiplication with any $a^{-}$ and elimination of the latter (by demanding that
$\widehat{\partial}_{x^{\alpha \, - \, \kappa}}(x^{\beta \, - \, \lambda}) = \delta^{\beta}_{\alpha} \delta_{\kappa}^{\lambda}$) fixes the hatted operators.  In the bosonic case, no such ambiguity arises.
Coordinate transformations due to quaternionic diffeomorphisms in the $x^{\alpha}, x^{\alpha \, \dag}$
lead to the usual tranformation rules:
$$\widehat{\partial}_{x^{\alpha \, - \, \kappa}} =
\frac{\widehat{\partial} x'^{\beta \, \pm \, \gamma}}{\widehat{\partial} x^{\alpha \, - \, \kappa}}
\widehat{\partial}_{ x'^{\beta \, \pm \, \gamma}}
+ \frac{\widehat{\partial} x'^{\beta \, \pm \, \dag}_{\gamma}}{\widehat{\partial} x^{\alpha \, - \, \kappa}}\widehat{\partial}_{ x'^{\beta \, \pm \, \dag}_{\gamma}}$$
and likewise for the complex and $\dag$ conjugates.  For now, this is all we need.
\subsection{Atomistic Calculus.}  The content of this section deals with delicate observations regarding the meaning of
differential calculus which are usually ignored or forgotten.  For simplicity, we shall deal with calculus in one real variable
but it is obvious how it translates to the general case.  On the real line $R$, we can put a global coordinate $x$ and define the associated derivate $\partial_x$ which maps functions
$f:R \rightarrow R$ to the functions $\partial_x f$.  Usually, one interprets the derivative
 of $f$ in a point $p$ as the application of the
derived function on $x(p)$ while only the behavior of $f$ in an infinitesimal neighborhood is required.  That is, the derivative
$\partial_{x| p}$ shouldn't have anything to do with $\partial_{x |q}$; therefore,
introduce the functions $\widehat{x}(p)$ which equal
$x$ in the infinitesimal neighborhood around $p$ defined by the coordinate interval
$\left[ x(p) - \epsilon, x(p)  + \epsilon \right]$ where $\epsilon$ is a positive
infinitesimal.  One should pay attention that $\widehat{x}(p)$ really is to be considered as
 a function attached to $p$ while
$x(p)$ is just an ordinary number.  Actually, the latter functions solve the duality problem
$$\partial_{x| p} \widehat{x}(q) = \delta_{p}^{q}$$ where the latter equals unity iff $p=q$ and
zero otherwise.  Quantum mechanics, as
 we shall see uses this atomisation to the fullest extend; that is,
it is written explicitely in the language of differentials $\partial_{x| p}$ and atoms $\widehat{x}(q)$.  Indeed, wave functionals
are functions of an $\aleph_1$ number of variables
$$\Psi \equiv \Psi(\widehat{x}(q); q \in R)$$ but then in the appropriate number of dimensions of course.  In the following sections,
we shall speak about the quaternions in different ways: (a) as belonging to algebraic representations of the Lorenz group (b) as a Lie algebra.  We first
introduce classical massless fermions in the quaternion language and next formulate a
 fully covariant dynamical principle for free Fermi quantum field theory.
\subsection{Free Fermi field theory.}
In this section, we treat free Fermi field theory in the quaternion formalism and comment upon
the differences and similarities with the standard Weyl representation.  The reason why we go through this
excercise shall be clarified later on.  Let us start by rewriting old equations in a different
language : the Dirac equation knows about causality through the quaternion algebra.  Indeed,
$$D = q^{m} e^{\alpha}_{m} \nabla_{x^{\alpha}}$$ where $x^{\alpha} = t,x,y,z$ is the usual
Dirac operator written in an inertial coordinate
 system and gauge, that is $e^{\alpha}_{m} = \delta^{\alpha}_{m}$
 and the (spin) connection $\nabla_{\mu}$ reduces to the standard partial derivatives;
  $\widetilde{D} D$ is the
Klein Gordon operator.  Now, the quaternion algebra has at least two product structures,
the usual one which we denote by concatenation and the
abelian product
$$\mathbf{p}\cdot \mathbf{q} = \frac{1}{2}\left( \mathbf{p}\mathbf{q} +  \mathbf{q}\mathbf{p}\right).$$
Given the latter, there are now two ways of constructing massless particle theories, as a left (right) module or
bimodule; specifically $D \Psi = 0$ ($\Psi \overleftarrow{D} = 0$) or $D \Psi = \Psi \overleftarrow{D} = 0$.  We
shall treat both in full detail.
\subsubsection{The left (right) module free field theory.}
Let us look for plane wave solutions of the kind $$\Psi = e^{i (Et - \vec{p}.\vec{x})} \widetilde{\mathbf{r}}$$
where $\widetilde{\mathbf{r}} = r_{m}\widetilde{q}^{m}$.  Then $D \Psi = 0$ if and only if
$\mathbf{p}\widetilde{\mathbf{r}} = 0$ where $\mathbf{p} = E1 - p^k (iq^k)$.  The latter equation
can be decomposed into $- E \vec{r} - r_0 \vec{p} +
  i \vec{p} \times \vec{r} = 0$ and $Er_0 + \vec{p}\cdot \vec{r} = 0$.  A general complex
  solution is given by $\widetilde{\mathbf{r}} = \alpha \widetilde{\mathbf{p}} + \widetilde{\vec{n}}$
  where $\vec{n}$ is some nonzero complex vector\footnote{Depending upon the sign of $E$ we have the right or left handed polarization vector of the photon.
  We conjecture here that the \emph{real} polarization vectors of the photon arise by taking squares of right and left handed Fermi particles.}
  satisfying $\vec{n}\cdot \vec{n} = \vec{n} \cdot \vec{p}= 0$
  and $\mathbf{p}\widetilde{\mathbf{p}}= 0$.  Since
  $\mathbf{p} \widetilde{\vec{n}} = 0$ one has that $
  E \widetilde{\vec{n}} = p^k (iq^k) \widetilde{\vec{n}}$  meaning that our general complex solutions have
  unit helicity; the helicity operator for particles with four momentum $\mathbf{p}$ being
   given by $$S(\mathbf{p}) = \frac{i q^{k}p_k}{E}.$$ Actually, if one takes the viewpoint
   that plane wave solutions,  with non real eigenvalues corresponding to real multiples of the
   physical operators $i \partial_{x^{\alpha}}$ and $(iq^k)$, are not measurable; then
   the complex polarization given by $\vec{n}$ is a non-measurable hidden variable and
   particles with complex momenta are not measurable at all.  The former statement is obvious since any
   vector $\vec{s}$ satisfying $\vec{s}\widetilde{r} = 0$ (with $\vec{r}\cdot\vec{r} =
   \alpha^2 E^2$
    and $(r_0)^2 = \vec{r}\cdot \vec{r}$) has to satisfy $\vec{s}\cdot \vec{r}=0$ and
   $r_0 \vec{s} - i \vec{s} \times \vec{r} = 0$ and all nonzero solutions to the latter equations are
complex.  Therefore, for real momenta $\mathbf{p}$, the only real operators of the kind
$s^k (iq^k)$ so that $s^k (iq^k) \widetilde{\mathbf{r}} = \lambda  \widetilde{\mathbf{r}}$ for
some $\lambda \in R$ are real multiples of the helicity operator.  So, even though hidden particles and real
particles with hidden properties exist, it appears that the predictive content of this theory
is exactly the same as it is in the Weyl representation.  We finish this section by studying
the issue of Lorentz invariance and quaternion valued action principles.  Define
$$\Lambda_{\frac{1}{2}} := e^{ \frac{1}{2}a_k (iq^{k})}$$ where $k:1 \ldots 3$ and
$a_k \in C$.  Letting $a_k = b_{k} + i c_k$ then it is easy to verify that $$\Lambda^{\dag}_{\frac{1}{2}}
q^{m} \Lambda_{\frac{1}{2}} = \Lambda^{m}_{n}\, q^{n}$$ where $$\Lambda = e^{b_k B^{k} + c_k R^k}$$
with the boost and rotation matrices given by $(B^k)^{m}_n = \delta^{m}_0 \delta^{k}_n +
\delta^{mk} \delta_{n0}$ and $(R^k)^{m}_n = \epsilon_{0 m n k}$.  Lorentz covariance now
 means that any solution
$$\Psi(\Lambda x')
=  \Lambda_{\frac{1}{2}} \Psi'(x')$$ where
$x = \Lambda x'$.  Indeed, $$D \Psi(x) = 0$$ if and only if
\begin{eqnarray*}
0 & = & q^m e_{m}^{\alpha} (\Lambda^{-1})_{\alpha}^{\beta} \partial'_{\beta} \Psi(\Lambda x') \\
& = & q^s (\Lambda^{-1})^{m}_s \Lambda_{m}^n \delta_{n}^{\alpha} (\Lambda^{-1})_{\alpha}^{\beta} \partial'_{\beta} \Psi(\Lambda x') \\
& = & (\Lambda^{-1})_{\frac{1}{2}}^{\dag}q^n (\Lambda^{-1})_{\frac{1}{2}} \delta_{n}^{\beta} \partial'_{\beta}\Psi(\Lambda x')\\
\end{eqnarray*}
and this last identity vanishes if and only if
$$q^n \delta_{n}^{\beta} \partial'_{\beta}\Psi'(x') = 0.$$  Hence, it is possible to define
quaternion valued Lorentz currents
$$j^{m}(x) = \Psi^{\dag}(x) q^m \Psi(x).$$  Similarly, the (scalar part of the) quaternion valued action
$$\mathcal{S}(\Psi, \Psi^{\dag}) = \int d^4x | \textrm{det}(e_m^{\alpha})|
\Psi^{\dag}(x) D \Psi (x)$$ is manifestly gauge and coordinate invariant.  We now proceed by
formulating the bi-module theory.
\subsubsection{The bimodule free field theory.}
The only naive worry of the left module free field theory concerns the existence of hidden complex
polarizations; there is a way to eliminate those however through considering the complex space
  $$\mathcal{W} = \{ \Psi | D \Psi = \Psi \overleftarrow{D} = 0 \, \textrm{where} \, \Psi \,
  \textrm{is complex quaternion valued}\}.$$   For plane wave solutions
  $$\Psi = e^{i (Et - \vec{p}.\vec{x})} \widetilde{\mathbf{r}}$$ this imposes the extra
  condition that $\vec{p} \times \vec{r} = 0$ leading to $\widetilde{\mathbf{r}} = \lambda \widetilde{\mathbf{p}}$.
  The unbounded solutions with complex momenta could be ignored as usual but I do not see any a priori
  reason for doing so.  Under a gauge transformation
  $\Psi$ transforms as $$\Psi(x) = \Lambda_{\frac{1}{2}} \Psi'(x) \Lambda_{\frac{1}{2}}^{\dag} \in \mathcal{W}$$ and
  performing a supplementary coordinate transformation $x = \Lambda x'$ takes the equations of motion into
  original form\footnote{It should be clear
that one cannot expect a unified theory to be Lorentz covariant in the above sense; from a
relativistic point of view
the coordinate transformations are meaningless since they do not represent isometries of the full dynamical metric anyway.}.
The Lorentz transform of
a plane wave $\Psi(x) = e^{ip_{\mu} x^{\mu}} \widetilde{\mathbf{p}}$ is given by
$$(\Lambda^{-1})_{\frac{1}{2}}\Psi(\Lambda x')(\Lambda^{-1})^{\dag}_{\frac{1}{2}} = e^{i(\Lambda_{\mu}^{\nu} p_{\nu})x'^{\mu}}
 (\Lambda^{n}_m p_n) \widetilde{q}^m$$ since $$\Lambda_{\frac{1}{2}}^{-1} \widetilde{q}^m (\Lambda_{\frac{1}{2}}^{-1})^{\dag}
 = \Lambda^{m}_{n} \widetilde{q}^n.$$  There is however another transformation law
 on $\mathcal{W}$.  Indeed,  the operator $q^m \delta_{m}^{\alpha} \partial_{\alpha}$ is invariant under the
 combined action of the quaternion conjugation and space reflection $R : (t, \vec{x}) \rightarrow
 (t, - \vec{x})$ implying the equation $D \Psi = 0$ is mapped to $$\Psi^R (x) \widetilde{D}_R
 = \widetilde{\Psi}(Rx)
  \widetilde{D}_R = 0.$$
  The Lorentz group action $\Lambda \rightarrow R \Lambda R$ obviously has the same effect:
  $$\Psi^R(x) \rightarrow \Lambda^{-1}_{\frac{1}{2}} \Psi^R(R\Lambda R x) (\Lambda^{-1}_{\frac{1}{2}})^{\dag}.$$
  It is instructive to explicitely stress the difference between the bi-module quaternion viewpoint and
  the standard group transformation laws.  In the latter, the Dirac operator transforms as
  -like the word says- an operator, while in the former it transforms in exactly the same way
  as the wave function does, that is as a (co) vector.  The quaternion product procures all the
  necessary operational properties in order to derive the traditional solution space spanned by
  right (left) handed particles as a quaternion bi-module.  This is not a minor point and it should
  be appreciated properly : everything we just did was \emph{intrinsic}; that is we didn't
   have to look for representation spaces of the quaternion algebra.  On the contrary, the quaternion algebra
   can be interpreted as what we shall call a Vecalg, that is a representation space and algebra
   at the same time; the Lorentz group has a natural mathematical significance within such structure.
   It is not difficult to understand that the possibilities of a Vecalg go way beyond standard
   representation theory; for example, it will allow us to derive the gauge charges of the
   standard model.  There is however a price to be paid, for example the standard Lorentz current
   $j^m(x) = \Psi^{\dag}(x) q^m \Psi(x)$ now transforms as $\Lambda^{m}_n \Lambda_{\frac{1}{2}}
    \Psi'^{\dag}(x) q^n \Psi'(x) \Lambda_{\frac{1}{2}}^{\dag}$ while the only acceptable
    transformation law would be of the kind $\Lambda^{m}_n \Lambda_{\frac{1}{2}} \ldots \Lambda_{\frac{1}{2}}^{-1}$ since
    the supplementary spin transformation preserves the algebraic properties of the original current
    four vector.  Hence, insisting upon an operator valued Lagrangian action principle not only requires
 new physics in the bi-quaternion formalism, but also implies that the latter is essentially non-hermitian.
 Most of these issues are undesirable and one must ask whether there is another way out; it turns out that an analogue
 of the Lorentz current can be found if and only if right and left handed particles belong to the same representation of the Lorentz group.  In other words,
 the bi-module theory \emph{appears} to assert that the Dirac picture is essentially the correct one.  Consider the action
 $$\mathcal{S}(\Psi, \Psi^{\dag}, \widetilde{\Theta}, \widetilde{\Theta}^{\dag}) =
 i \int d^4x | \textrm{det}(e_m^{\alpha})| \left(
                                           \begin{array}{cc}
                                             0 & \Psi^{\dag}(x) D \Psi(x) + \Psi(x) \overleftarrow{D} \Psi^{\dag}(x)  \\
                                            \widetilde{\Theta}^{\dag}(x) \widetilde{D}
                                            \widetilde{\Theta}(x) +
                                           \widetilde{\Theta}(x) \overleftarrow{\widetilde{D}}
                                           \widetilde{\Theta}^{\dag}(x)  & 0 \\
                                           \end{array}
                                         \right)
  $$ then it is easy to check that under a (combined) Lorentz transformation, the latter transforms as
  $$\mathcal{S}(\Psi, \Psi^{\dag}, \widetilde{\Theta}, \widetilde{\Theta}^{\dag}) =
  \left(
    \begin{array}{cc}
      \Lambda_{\frac{1}{2}} & 0 \\
      0 & (\Lambda_{\frac{1}{2}}^{\dag})^{-1}
        \\
    \end{array}
  \right) \mathcal{S}(\Psi', \Psi'^{\dag}, \widetilde{\Theta}', \widetilde{\Theta}'^{\dag})
  \left(
    \begin{array}{cc}
      \Lambda_{\frac{1}{2}}^{-1} & 0 \\
      0 & \Lambda_{\frac{1}{2}}^{\dag}
        \\
    \end{array}
  \right)$$  as it should.  Variation of the above action with respect to (say) $\Psi^{\dag}(x)$ gives
  $$\delta \Psi^{\dag}(x) D \Psi(x) + \Psi(x) \overleftarrow{D} \delta \Psi^{\dag}(x) = 0.$$
  We prove now that the latter is equivalent to $D \Psi = \Psi \overleftarrow{D} = 0$.  Indeed, the former
  implies\footnote{$\mathbf{p} \mathbf{q} + \mathbf{q}\mathbf{r} = 0$ for all $\mathbf{q}$ implies that
  $(p+r)q = 0$, $(\vec{p} + \vec{r})\cdot \vec{q} = 0$ and $i (\vec{p} - \vec{r}) \times \vec{q} = 0$ leading to
  $\vec{p}= \vec{r}= 0$ and $p + r = 0$.} that $D \Psi(x) = - \Psi(x) \overleftarrow{D} \in C$
  and since the scalar parts of both expressions are
  equal, they have to vanish. \\* \\  Although we have a conserved Lorentz
  current $j^m$ which transforms appropriately $$j^m(x) = \left(
                                           \begin{array}{cc}
                                             0 & \Psi^{\dag}(x) q^m \Psi(x) + \Psi(x) q^m \Psi^{\dag}(x)  \\
                                            \widetilde{\Theta}^{\dag}(x) \widetilde{q}^m
                                            \widetilde{\Theta}(x) +
                                           \widetilde{\Theta}(x) \widetilde{q}^m
                                           \widetilde{\Theta}^{\dag}(x)  & 0 \\
                                           \end{array}
                                         \right)$$ the only way to get a scalar out is to consider squared currents.
From the physical point of view, this is not absurd; indeed measurements correspond to current interactions
which doesn't
need to imply that the individual currents have nontrivial scalar parts.  Indeed,
it might be there is something deep about the vanishing of $Tr(j^m)$ in the light of background
independence.  Apart from the fact that $j^m$ is in general not the measured current,
it does not correspond to the center of mass current of the particle either (cfr. zitterbewegung); moreover,
one has a conserved current in the bi-modal quaternion formalism given by $\mathcal{S} \left( \Psi(x)
q^m \right)$.  For plane waves $\Psi(x) = \widetilde{p} e^{ip_{\mu}x^{\mu}}$, the Lorentz current in the
left modal formalism is given by $2 p_0 p^m$ which doesn't look very covariant
because of the factor $p_0$ (a similar unnatural normalization occurs in the Weyl representation) while
the covariant current in the bi-modal case is given by $p^m e^{ip_{\mu}x^{\mu}}$. \\* \\
\underline{Remark} \\ Although the above equations are linear, it is rather easy to see
 that all solutions $\Psi$ of $D \Psi = 0$ also satisfy
 $$ V^{\kappa} \partial_{x^{\kappa}} V^{\alpha} = 0$$ where $$ V^{\kappa} = \Psi q^{\kappa}.$$
  The latter is an ordinary geodesic equation for quaternion valued vector fields which brings us in closer correspondence with gravity.\\* \\
  \underline{Conclusions} \\  One disposes now of at least four different ways,
   from which three are inequivalent, to look at classical free Fermi field theory: (a) in the Dirac
   representation (b) the Weyl spinor representation (c) the uni modal and (d) bi-modal quaternion theories.
   \subsection{A remark about Mass and inclusion of Gauge Theories.}
   Mass cannot be introduced in the bi-quaternion theory in the same way as it happens in the Weyl representation
   or uni-modal theory since it should effectively transform as
   $$m = \Lambda_{\frac{1}{2}} m' \Lambda_{\frac{1}{2}}^{\dag}$$ at least when one insists upon keeping
   the opposite chiral field in the coupling.  What does the above mean?  Does it indicate that the bi-modal
   quaternion theory is a lost case or on the contrary that it reflects that mass is due to self interaction?
   The above transformation property reveals the non-trivial point that - since real measured
   currents are interactions of
   $\Psi$ and $\widetilde{\Theta}$ in the bi-modal quaternion formalism (no need for that in the uni-modal and Weyl representation) -
   the real measured mass would behave like $\widetilde{m}m$; so there is no obvious problem
    concerning this viewpoint.  The latter implies of course that we should be able to escape
     from the usual renormalization problems, it is
    my hope that the non abelian nature of the coordinate space would solve that problem. \\* \\  A related
    discussion deals with why the gauge coupling constants are what they are; given the formulation of gauge
    theory, one cannot escape the fact that these constants are free.  Perhaps,
    one has to reformulate
    gauge theory so that an absolute scale arises (as is the case for gravity) or
    perhaps self interactions lead to effective values which depend upon the algebraic
     properties of the gauge algebra's and initial conditions of the gauge fields only.  We
     argue now why the first option has to be taken seriously (for further reasons) and present an
     alternative.  For simplicity we discuss Maxwell theory and at the risk of appearing too negative, let us start by
     summing up the shortcomings: \begin{itemize} \item The conservation law for the Lorentz current
     $\partial_{\mu} j^{\mu} = 0$ appears to be incorrect if such current gives rise to radiation (this is another form of
     the famous self interaction problem as we shall see shortly).
     \item The gauge transformation law $A_{\mu} \rightarrow A_{\mu} + i \partial_{\mu} \Omega$ under the action of a local
     $U(1)$ transformation $e^{i \Omega}$ should only hold on the support of the Fermi field $\Psi$; there is no reason why
     it should extend into the vacuum region.
     \item One has undesirable vacuum solutions.
     \item A manifestly causal representation incoorporating the gauge transformation laws for
     $A_{\mu}$ is lacking.  In other words, the action should be manifestly gauge invariant
     and not only when the classical equations of motion are satisfied.
     \item There should be no free scales.
    \end{itemize}
    We show now that there exists a solution for all these problems and the result is a theory which is
    distinctly different from classical Maxwell electrodynamics.  Let us start by constructing such solution, show it satisfies all above criteria
    and then say something about uniqueness (we work in the quaternion representation).  It is by no means the intention of this note to prove that
    everything works out as it should; we merely aim to open a
    new avenue for gauge theory which is in line with the global picture developped in this paper.  Let us start by considering a partially wrong solution\footnote{What follows below only works naturally in the quaternion formalism
   since it is of utmost importance that $\widetilde{\Theta}$ and $\Psi$ have the same electric charge.  In the Weyl
    representation, one would need a $\widetilde{\Theta}^{\dag}$ in order to get the right transformation properties under the spin group
    which comes at the cost of opposite charges.  One might think that by not considering the
     Dirac spinor as a single particle and by attributing opposite charges to the left and
    right handed components (by use of the $\gamma^5$ matrix), the Weyl representation is saved.  Albeit this is indeed the case for the following
     incomplete proposal; the correct formula
    for the gauge potential below requires $\Psi \widetilde{\Theta}$ to be invertible which cannot be achieved in the Weyl representation
    since $\left( \Psi_{R} \right)^a \left( \Psi_{L}^{\dag} \right)_b$ has at most rank one.  This, again, could be corrected by demanding the existence of at least two different Dirac spinors,
    but one might consider such option not very natural given that one would reasonably expect one field to (largely) dominate at some spatial locations giving rise to nearly singular matrices.}:
    $$A_{\mu}(x) =  \frac{1}{8 \pi} \int d^4y \, \sqrt{- g(y)} \partial_{\mu}^x G_{R} (y,x) $$ $$\frac{\mathcal{S}(\widetilde{\Theta}(y)
    g^{\alpha \beta}(y) \nabla_{\alpha} \nabla_{\beta} \Psi (y))
   + \mathcal{S}(\widetilde{\Theta}(y) g^{\alpha \beta}(y) \overleftarrow{\nabla}_{\alpha} \overleftarrow{\nabla}_{\beta} \Psi (y))
   - 2 \mathcal{S}(\widetilde{\Theta}(y) g^{\alpha \beta}(y) \overleftarrow{\nabla}_{\alpha} \nabla_{\beta} \Psi (y))
    }{\mathcal{S}(\widetilde{\Theta}(y) \Psi (y))}$$ where $\mathcal{S}(\cdot)$ denotes the scalar part.
    First, some comments are in place: an electromagnetic
    field is entirely generated by the interaction of left and right handed massless fermions,
     the latter which is known to give mass to fermions.  Moreover, the entire dynamics is invariant under rescaling of
    the Fermi fields given that $A_{\mu}$ only depends upon the wave vectors, not upon the amplitudes.  This is exactly the kind of scaling invariance
    which we meet in quantum mechanics.  Under
    $\Psi \rightarrow e^{i \Omega} \Psi$ and $\widetilde{\Theta} \rightarrow e^{i \Omega} \widetilde{\Theta}$, $A_{\mu}$ transforms as it should since:
    \begin{eqnarray*} A'_{\mu}(x) & = & A_{\mu}(x) + \frac{i}{4 \pi} \int d^4y \, \sqrt{- g(y)} \partial_{\mu}^x G_{R} (y,x) g^{\alpha \beta}(y) \nabla_{\alpha} \nabla_{\beta} \Omega(y) \\
    & = & A_{\mu}(x) + i \int d^4y \partial_{\mu}^x \delta^{(4)}(x - y) \Omega(y)  \\
    & = & A_{\mu}(x) + i \partial_{\mu} \Omega(x)
    \end{eqnarray*}  The latter transformation law only holds on the intersection of the
     supports of both left and right handed partners (which means they should be identical)
     and such details are postponed to a future paper.  In the vacuum region, a similar computation yields that
    $$g^{\alpha \beta}(x) \nabla_{\alpha} \nabla_{\beta} A_{\mu}(x) = 0$$ and
    $$g^{\alpha \beta}(x) \nabla_{\alpha} A_{\beta}(x) = 0.$$
    \\ That is, what one usually calls the Lorentz gauge gives nothing but the real field potential.  All the above results only hold
    for flat spacetime, a general expression for curved spacetime would yield curvature corrections.  Clearly, there are no vacuum solutions for the electromagnetic field as well as no
    free coupling constants (if fields of different charges are added, one notes that only the relative charges count).  Let us further comment, since
    $A_{\mu}(x) = \partial_{\mu} K(x)$, it is easy to see that in strict Fermi vacua,
    the above gauge field consists of longitudonal polarizations only; that is, only the Coulomb field is included for now and not the
    radiation  modes.  Now, the only way to recover them is to consider a full quaternion valued expression such as
    $$A_{\mu}(x) =  \frac{1}{8 \pi} \int d^4y \, \sqrt{- g(y)} \partial_{\mu}^x G_{R} (y,x)
    \left( \Psi(y) \widetilde{\Theta} (y) \right)^{-1}$$ $$\left( \Psi(y)
    g^{\alpha \beta}(y) \nabla_{\alpha} \nabla_{\beta} \widetilde{\Theta} (y))
   + \Psi(y) g^{\alpha \beta}(y) \overleftarrow{\nabla}_{\alpha} \overleftarrow{\nabla}_{\beta} \widetilde{\Theta}(y)
   - 2 \Psi(y) g^{\alpha \beta}(y) \overleftarrow{\nabla}_{\alpha} \nabla_{\beta} \widetilde{\Theta} (y) \right).$$  Indeed, the (quantum mechanical)
   spin is in the quaternion algebra and $A_{\mu}$ transforms in the adjoint representation of the spin group as a genuine spin one particle should\footnote{The lack of such representation in
   the standard Yang-Mills formulation is worrysome to say the least.}.  $A_{\mu}$ acts from the left on
   $\Psi$ and from the right upon $\widetilde{\Theta}$; since in the bimodule quaternion formalism, no distinction
   between left and right exists, it would seem natural to consider the field potential $B_{\mu}$
   which is defined from $A_{\mu}$ by switching the positions of $\Psi$ and $\widetilde{\Theta}$.
    We argue now in the uni-modal formalism how tranversally polarized waves are
   constructed and how this is connected to the process of mass and spin generation for the fermions.
   Considering the coordinate and gauge invariant (with respect to Lorentz and local $U(1)$ transformations)
    action principle given by
   $$\mathcal{S} = i \int d^4 x \sqrt{- g(x)} \mathcal{S} \, [ \Psi^{\dag}(x) e^{\mu}_m q^m \left( \nabla_{\mu}
    - A_{\mu}(x) \right) \Psi(x) - \Psi^{\dag}(x) \left( \overleftarrow{\nabla}_{\mu}
    - A^{\dag}_{\mu}(x) \right) e^{\mu}_m q^m  \Psi(x) $$ $$ + \, \widetilde{\Theta}(x) \left( \overleftarrow{\nabla}_{\mu}
    - A_{\mu}(x) \right) e^{\mu}_m \widetilde{q}^m  \widetilde{\Theta}^{\dag}(x) -
    \widetilde{\Theta}(x)  e^{\mu}_m \widetilde{q}^m  \left( \nabla_{\mu}
    - A_{\mu}^{\dag}(x) \right) \widetilde{\Theta}^{\dag}(x) ]$$ then variation with respect
    to $\Psi^{\dag}(x)$ gives in an inertial coordinate system
    $$\mathcal{S} \left( \delta \Psi^{\dag}(x) e^{\mu}_m \left(  2 q^m \partial_{\mu}  - q^m A_{\mu}(x) + A^{\dag}_{\mu}(x) q^m
    \right) \Psi(x) \right) \, +  $$ $$\frac{1}{8 \pi} \mathcal{S} \left( \int d^4 y \Psi^{\dag}(y) \partial_{\mu}^y G_R(x,y)
    \delta_{\Psi^{\dag}(x)} B(x) e^{\mu}_m q^m  \Psi(y) \right) \, + \,$$ $$ \frac{1}{8 \pi} \mathcal{S} \left(
    \int d^4 y \widetilde{\Theta}(y)
    e^{\mu}_m \widetilde{q}^m  \partial_{\mu}^y G_R(x,y)
    \delta_{\Psi^{\dag}(x)} B(x)   \widetilde{\Theta}^{\dag}(y)\right) $$ where $B(x)$ equals
    $$\left( \widetilde{\Theta}^{\dag} (x)
    g^{\alpha \beta}(x) \nabla_{\alpha} \nabla_{\beta} \Psi^{\dag}(x)
   + \widetilde{\Theta}^{\dag}(x) g^{\alpha \beta}(x) \overleftarrow{\nabla}_{\alpha}
   \overleftarrow{\nabla}_{\beta} \Psi^{\dag}(x)
   - 2 \widetilde{\Theta}^{\dag}(x) g^{\alpha \beta}(x) \overleftarrow{\nabla}_{\alpha} \nabla_{\beta} \Psi^{\dag}(x) \right)$$ $$\left( \widetilde{\Theta}^{\dag} (x) \Psi^{\dag}(x) \right)^{-1}.$$
   The resulting equations of motion are fairly elaborate and, at this point, it is much more
   instructive to comment upon their structure.  One notices that mass corrections arise
    due to electromagnetic radiation being sent out to the future.  That is, the \emph{incoming} radiation is represented by
    an integral over the past lightcone and the latter minimally couples to the spinor,
    the \emph{mass corrections} come from additional terms
    evaluated in the same space time point and finally, the \emph{outgoing} radiation
    is represented by an integral over the future
    lightcone which couples to the spinor of opposite chirality.
       These correction terms are entirely
    absent in the standard formulation while their presence should be obvious from the
    physical point of view\footnote{Even worse, the standard formulation
    does not contain past, nor future.}.  It is clear that the above construction is not unique because the two factors determining
    $A_{\mu}$ do not commute; it appears to me that it is worthwhile to investigate proposals of the
    above type in full detail and see where they lead us.
    \\* \\ Before going over to the quantum theory, some further comments are needed.
\subsubsection{What is quantum theory?  A simple example.}
Here, we make some further remarks upon the operator valued action principles defined previously
and jump ahead into the next section by making specific comments upon the (existence of the) quantum field theoretical
formulation.  Let us start by mentioning that from the extremal variation point of view, there is no
problem whatsoever with the left and bi-modal action principles.  However, from the quantum mechanical
point of view, further problems arise.  At first sight, the left-modal action poses no problem
 since it is Hermitian and one can restrict to the unit quaternion component without breaking
 Lorentz invariance; the bi-modal action however is not hermitian and its scalar part is trivially
  zero.  We do not treat the problem of quantization in the quaternion formalism in this paper,
  but we concentrate instead upon suitably reformulating (free) quantum field
 theory\footnote{A more precise formulation as well as philosophical discussion
  is given in the next section}.
  The latter is defined by a non trivial two-point specification $A(\cdot , \cdot)$ of the spacetime operator field
   algebra defined amongst the objects $\Psi_m(x)$, $\Psi^{\dag}_n(y)$, respecting (a) the laws of causality with time
   orientation, (b) symmetric in the
  spacetime points $x,y$ and (c) closed with respect to hermitian conjugation.  All this implies that
  $A(\alpha, \beta) = a \left( \alpha \beta \pm \beta \alpha \right)$ where $a$ is a real number.  The laws of causality are
   represented by the radiative Green's function $\mathcal{G}_A(x,y)$ and the operators
   $D, \widetilde{D}$ respectively since those are the only gauge and coordinate invariant
   objects (the former with distributional support on lightcones) trivially formed by causality.
   Moreover, demanding the basic algebra to be first
   order in $D, \widetilde{D}$ singles out the anti-commutator (one could see this as another form
   of the spin statistics theorem).  Indeed, there are two possible expressions\footnote{Of course $D$ and $\widetilde{D}$ could be exchanged.} respecting the above symmetries,
   that is the symmetric one $$\lambda \left( D_x - D_y \right) \mathcal{G}_A(x,y)$$ and the anti-symmetric one
   $$\lambda \left( D_x + D_y \right) \mathcal{G}_A(x,y).$$  A small computation reveals that the latter vanishes in flat spacetime
   and therefore $A(\alpha, \beta) = a \left( \alpha \beta + \beta \alpha \right)$.  For right handed (that
   is, we study the theory attached to $D$) Weyl
    spinors $\Psi^{j}$,
   $j=1,2$, this leads to the laws
   $$ \{ \Psi^j(x), \Psi^k(y) \}  = 0$$
   and $$\{ \Psi^j (x), \Psi_{k}^{\dag}(y) \} = \mu \left( \widetilde{D}^{j}_{k \, x} - \widetilde{D}^{j}_{k \, y} \right) \mathcal{G}_A(x,y)$$ with
   $\mu \in R$.  Indeed, the former expression has to vanish since the only nontrivial causal expressions transform as
   $\Lambda_{\frac{1}{2}}^{-1} \ldots \left( \Lambda_{\frac{1}{2}}^{-1} \right)^{\dag}$.  Now,
   the only important property is the sign of $\mu$ since any absolute value can be adsorbed into the definition of the
   field operators.  The above is correct if and only if global gauge transformations are allowed;
   under local gauge transformations
   the second equality is not even well defined\footnote{In the Weyl formalism, a spinor valued
   connection $\Gamma^{j}_{\mu k}$ would be
   required for this.}.  \\* \\
   We now derive the usual Fock algebra for intertial observers
   given by a foliation $(t,\vec{x})$.  Choosing a time $t_0$ and letting $\delta > 0$, we denote by
   $\Sigma^{\delta}_{t_0}$ the tickened hypersurface $\left[ t_0 - \delta, t_{0} + \delta \right] \times R^{3}$.
   Within $\Sigma^{\delta}_{t_0}$, we define the algebra's $\mathcal{A}^{s}$ generated by
   smeared operators (and their conjugates) of the kind $$\int_{R^3} d\vec{x} f(\vec{x}) \Psi^j (s,\vec{x})$$ for
   $t_0 - \delta < s < t_0 + \delta$ and $f \in \mathcal{S}(R^{3})$, the Schwartz functions on space.
   The interpretation of the spacetime algebra is formed by identifying the algebra's $\mathcal{A}^{s}$ with
   $\mathcal{A}^{t_0}$ using the spacetime algebra
   $$\{ \Psi^j (t, \vec{x}), \Psi^k (s, \vec{y}) \} = 0$$ and
   $$\{ \Psi^j (t, \vec{x}), \Psi_{k}^{\dag}(s, \vec{y}) \} = \left( \widetilde{\sigma}^{\mu} \right)^{j}_k
   \int \frac{d \vec{p}}{2(2 \pi)^{3} |\vec{p}|}
    p_{\mu} \left[ e^{i p(x - y)} + e^{i p(y - x)} \right].$$  This effective breaking of manifest gauge invariance
    uniquely results into
    $$\Psi^{\dag}_j (s, \vec{y}) = \left( \widetilde{\sigma}^{\mu} \right)^{k}_j \int
    \frac{d \vec{p}}{2(2 \pi)^{3}|\vec{p}|}
    \int d \vec{z} \, p_{\mu} \left[  e^{i p(z - y)} + e^{i p(y - z)} \right] \Psi^{\dag}_k (t_0, \vec{z})$$ and
    Fourier decomposition
    $$\Psi^{\dag}_j (s,\vec{y}) = \int \frac{d \vec{p}}{2(2 \pi)^{3}|\vec{p}|}\,
    e^{i \vec{p}.\vec{y}} \, a^{\dag}_j (\vec{p},s)$$ leads therefore to the relationship
    $$a^{\dag}_j (\vec{p},s) = \frac{1}{2|\vec{p}|} p_{\mu}
    \left[ \left(\widetilde{\sigma}^{\mu}\right)^{k}_j e^{i|\vec{p}|(t-s)}
    + \left( \sigma^{\mu} \right)^{k}_j e^{i|\vec{p}|(s-t)} \right] a^{\dag}_k (\vec{p},t)$$ for all $s,t \in \left( t_0 - \delta, t_0 + \delta
    \right)$.   Hence,
    $$a^{\dag}_j (\vec{p},s) =  n_j e^{i|\vec{p}|(t_0 - s)} b^{\dag}(\vec{p},t_{0})
    + m_j e^{i|\vec{p}|(s - t_0)} c(\vec{p},t_0)$$ where $p_{\mu} \left( \widetilde{\sigma}^{\mu}
    \right)^{k}_j n_k = 2 |\vec{p}| n_j$, $p_{\mu} \left( \sigma^{\mu}
    \right)^{k}_j n_k = 0$ and the opposite for $m_j$.  Interpreting $i \partial_s$ as the energy operator
    allows one to interpret $e^{i|\vec{p}|(t_0 - s)} b^{\dag}(\vec{p},t_{0})$ as the creation operator of a
     quantum with energy $|\vec{p}|$; and likewise $e^{i|\vec{p}|(s - t_0)} c(\vec{p},t_0)$ as the annihilation of a
     quantum with the same energy. \\* \\ We just formulated free quantum field theory in a manifestly Lorentz and coordinate invariant
     manner starting from four simple postulates.  Subsequently, we have shown how the former can
     be turned into the standard formulation by introducing an observer.  It is worthwile to
     note that we have no trouble with infinities arising
     in the Hamiltonian while quantizing the classical theory; the latter is constructed
      a posteriori, that is after the observer has been chosen.  We continue to formalize all this further on in the next section.
\subsection{Free Quantum Field theory.}
In this section, we go with free quantum field theory in two opposite directions.  The first one
takes QFT as it stands and reformulates it from first principles in a manifest gauge and coordinate
invariant manner; this road will lead us further apart from classical physics as we know it.  We shall explicitely
define the meaning of background independence and how to include interactions in this context.  The second road
 involves a geometrical representation of QFT on the bi-quaternion grasmannian geometry developped
 previously.  The way in which the representation works indicates that QFT is close to a non abelian
 statistical mechanics of a non-abelian field theory on extended space-time.  The latter
 constitutes a direction for local realism in which Bell's theorem is surpassed by going over to
 more complicated space-time structures.  Concerning the first road, it is worthwhile to notice that
 (a) we do not start from a classical world which we
 quantize either in the canonical or path integral way (b) dynamics is governed by
the demand of causality whereas in the ordinary construction this is usually a result.
Moreover, the formulation of the dynamics below does not require a
non-dynamical foliation having some physical significance, neither an initial nor
final hypersurface having a similar meaning.  Indeed, a theory is only background independent
if all physical predictions are composed of dynamical objects only.  Quantum mechanics is
in my opion not such theory and therefore diffeomorphism invariance has to be broken at
some point.  The only relevant question, if quantum mechanics itself is not to be modified,
is whether the essential operator content of quantum field theory can be given a manifest
background independent formulation.  The answer to this issue is a resounding yes and we
shall comment how this can be done.
\subsubsection{Manifestly background independent QFT.} First, let us start by making some
 comments upon the formulation of standard quantum field theory which are
required to comprehend what is going on next.  Although everyone with an elementary quantum
field theory background will know the first part of what I am going to say,
 the conclusions drawn in the second part are appearantly not frequently made.  The latter
 constitutes a radical departure from the traditional formulation; not only are the laws
we arrive at more simple and elegant, they moreover reflect an issue concerning the
mathematical formulation of the theory.  The latter concerns the true degrees of freedom
required to formulate the theory as well as the appropriate choice
of representation of the causal structure.  In order to grasp the former, let us work in the
canonical Heisenberg picture and denote for simplicity with $H$ the time independent
 Hamiltonian and $U(t) = e^{iHt}$ the unitary time evolution operator.  Now, in free quantum field theory, all local operators are
constructed from $\Psi(x)$, $\overline{\Psi}(x)$ and states are formed by applying
such smeared-out operators to the vacuum state $|0>$.  Now, anyone knows that the latter
objects are only determined upon unitary conjugacy (such as time evolution) and that making calculations requires one to pick out a specific
representation.  However, this really means that there is no quantum dynamics!  Indeed, ordinary quantum field theory in the
Heisenberg picture is formulated as being first order in time but the particular expressions attached to the initial operators
$\overline{\Psi}(0, \vec{x})$ and $\Psi(0, \vec{x})$ do not matter, only
their anti-commutator algebra is important and the latter is enforced \emph{a priori}.
Therefore, the equations fixing the anti-commutator algebra cannot contain derivatives of the
field operators but only the operators themselves.  The sublety here is that one cannot
formulate a theory by thinking in terms of a representation which would indeed require
specific intial operators, but in terms of the fundamental physical variables themselves.
A representation should always come in the end and never at the beginning.  That is, field
operators are what one could call objective degrees of freedom; the complex structure
distinguishing particles from anti-particles as well as the associated
  quantum states are a matter of subjective interpretation.  They have no a priori place in
  the dynamical formulation of the theory at all, but they sneak in at the end when
  interpretations in terms of measurements are made.  Indeed, using this dichotomy operator
   algebra/state to the
  fullest extend allows one to have a truely covariant formulation of the constraints the
  field operators have to satisfy.  Of course, such weltanschaung is strange to say the least,
   but a few useful lessons can be drawn nevertheless: (a) for those who think quantum theory
   is absolute, it provides a satisfactory starting point to include gravitational degrees of
   freedom (b) for the realists, it is the geometrical representation on the manifold itself
   which is interesting in trying to formulate generalized local realist theories.
   In other words, quantum theory as thus reformulated can be reconciled with general
   relativity.  \\* \\  We now arrive at the second issue, that
  is how to implement causality when no derivatives of the fundamental field operators are
  allowed for.  Also here, we will show that issues such as representations of the Lorentz
   group do not enter the formulation of the theory and that the latter is defined in a fully
    covariant way albeit it is of course beneficial to calculate
in an inertial coordinate system.  The notion of a dynamical principle should be replaced by
a causal Markov property on the fermionic spacetime field anti-commutator algebra.
During my investigations, I learned that not the same but a similar
approach towards axiomatic QFT is under construction by Hollands and Wald which appears
worthwhile to follow\footnote{Unfortunately,
this author did not find the time to study their construction.}.
We now sketch how to work out this programme by formulating some first principles.  \\
 \newtheorem{Principle}{Principle}
 \begin{Principle}
 As elucidated in the previous section; a quantum theory in the Weyl representation can be defined
 using the axioms of (a) space-time symmetry in the points $x,y$ (b) the algebra is closed with respect to
 Hermitian conjugation (c) causality is implemented using the Green's function $G_{A}(x,y)
 = G_{R}(x,y) - G_{R}(y,x)$ and (d) the algebra relations contain only first order derivatives.
\end{Principle}
As commented in the previous section, we derive the standard Fock representation by choosing a
 foliation, identify the algebra's associated to the $3$-spaces and derive the
 annihilation/creation operators through Fourier decomposition.  Given a notion of space and time,
 it is possible to define the Hamiltonian (up to an additive shift) through the standard Heisenberg
  relation.  The existence of an Hamiltonian is not trivial by any means and it should follow
  from the structure of the algebra itself: a necessary and sufficient condition for the
  latter to exist is expected to be
  equivalent to a generalization of the following Markov property : for all $x$ spacelike or past timelike related to $z$ and inertial spacelike hypersurface $\Sigma$ between $x$ and $z$,
  the following property
 $$\frac{1}{2} \left( \widetilde{D}_x - \widetilde{D}_z \right)
   \mathcal{G}_R(x,z) = \frac{1}{4} \int_{\Sigma} d\vec{y} \sqrt{h}
   \left( \widetilde{D}_x - \widetilde{D}_y \right)
   \mathcal{G}_R(x,y)\cdot \left( \widetilde{D}_y - \widetilde{D}_z \right)
   \mathcal{G}_R(y,z)$$ holds.  Implementation of the gravitational field implies that the
   Green's function itself is dynamical and the algebra should be enriched with two-point
   relations defined amongst the metric components and spinor fields and the metric components separately.
\subsubsection{A hint for extended local realism.}
The aim of this subsection is to represent standard free quantum field theory on the tangent bundle of
our generalized geometrical setting.  Let me stress that what follows should be an
 approximation to the full dynamical theory and is by no means the fundamental theory itself.
   Assume the global space-time manifold $\mathcal{M}$ is given by
    $R^{4} \oplus \mathcal{I}$ where for any open $\mathcal{O}$ in
    $R^{4}$, $\mathcal{O} \oplus \mathcal{I}$ corresponds to an infinitesimal strip in $\mathcal{S}$.  More specifically,
    the coordinates $x^{\alpha}$ are given by $$x^{\alpha} =  \left( a^{\alpha \, 0} \, + \, i \epsilon^{\alpha}  \right) \left( 1_L + 1_R \right) \,  + \,
   \left( \delta^{\alpha \, 0} \, + \, i \epsilon^{\alpha \, 0} \right) \left( 1_L - 1_R \right) \, + $$ $$\sum_{H \in \{ L,R \} } \left((\delta^{\alpha \, j}_H + i \epsilon^{\alpha \, j}_H)
     q_{j \, H} + (\delta^{\alpha s \, \kappa}_H + i \epsilon^{\alpha s \, \kappa}_H )q_{\kappa \, H} n_s \right)$$
    where the $n_s$ span the nilpotent part of $\mathcal{N}$, and the $\epsilon, \delta$ are
    infinitesimal real numbers or zero.
Given an inertial coordinate system $x^{(\upsilon) \, \alpha \, \kappa}_J, {x^{(\tau) \, \beta}_{\kappa \, K}}^{\dag}$,
we have the usual foliation of the abelian coordinates given by
 $$ \sum_{H \in \{ L,R \}} \frac{1}{2} \left( tr(x^{\alpha \, + \, 0}_H) + tr({x^{\alpha \, +}_{0 \, H}}^{\dag}) \right)$$
where $\upsilon, \tau$ denotes one of the four involutions : $\pm$ and the complex conjugate
of $\pm$.  By definition $\sum_{H \in \{ L,R  \} } \frac{1}{2} \left( tr(x^{0 \, + \, 0}_H) + tr({x^{0 \, +}_{0 \, H}}^{\dag}) \right)$ is the time coordinate and $\sum_{H \in \{ L,R  \}} \frac{1}{2} \left( tr(x^{j \, + \, 0}_{H}) + tr({x^{j \, +}_{0 \, H}}^{\dag}) \right)$, $j = 1 \ldots 3$, the usual space coordinates.  Denote by $\mathcal{R}_n$
a series of regular rasters satisfying $\mathcal{R}_n \subset \mathcal{R}_m$ for $n < m$ and
the limit raster $\mathcal{R}_{\infty}$ is dense in $\mathcal{M}$; furthermore, let $\mathcal{P}^{T}_n$ be the horizontal strip
$\left( R^{3} \times \left[ - T, T \right] \right) \oplus \mathcal{I} \cap \mathcal{R}_n$.  Each raster point $x^{i}_{n} \in \mathcal{R}_n$ is the midpoint of a cell
with volume $\left( \frac{1}{n} \right)^4$ and the latter partition $\mathcal{M}$.  Attached to $\mathcal{R}_n$ is the tensor algebra obtained by treating the coordinates of $x^{i}_{n}$
as independent variables which can undergo infinitesimal displacements.  In other words,
functions $\Psi$ look like $$\Psi \left( x^{i \, (\upsilon) \, \alpha}_{n \, H} ,{x^{j \, (\tau) \, \beta}_{n \, K}}^{\dag} \right)  $$ and we proceed now
by defining the appropriate creation and annihilation operators.  The operators we
have in mind now are of the kind $$a^{i \, (k)}_{n \, \alpha \, \kappa \, H} =
\widehat{\partial}^{(k)}_{x_{n \, H}^{i \, (-) \, \alpha\, \kappa}} + x^{i \, (-) \, \alpha \, \dag}_{n \,  \kappa \, H}$$
and $$\left( a^{i \, (k) \, \kappa}_{n \, \alpha \, H} \right)^{\dag} =
\widehat{\partial}^{(k)}_{ x^{i \, (-) \, \alpha \, \dag}_{n \, \kappa \, H}} + x^{i \, (-) \, \alpha \, \kappa}_{n \, H}$$
as well as their complex conjugates $$\overline{a}^{i \, (k)}_{n \, \alpha \, \kappa \, H} =
\widehat{\partial}^{(k)}_{ \overline{x_{n \, H}^{i \, (-) \, \alpha \, \kappa}}} +
\overline{x^{i \, (-) \, \alpha}_{n \, \kappa \, H}}^{\dag}$$ and
$$\left( \overline{a}^{i \,(k) \, \kappa}_{n \, \alpha \, H} \right)^{\dag} =
\widehat{\partial}^{(k)}_{\overline{x^{i \, (-) \, \alpha}_{n \, \kappa \, H}}^{\dag}} +
\overline{x^{i \, (-) \, \alpha \, \kappa}_{n \, H}}.$$
As usual, the index $k$ denotes that the operator acts upon
$W^{(k-1)}\left( \{ \omega^{(l)} \}_{l \leq k-1} \right)$.  One can easily verify that
$$\{ a^{i}_{n \, \alpha \kappa \, H} , a^{j}_{n \, \beta \, \lambda \, K} \} \Psi = 0
= \{ \overline{a}^{i}_{n \, \alpha \kappa \, H} , a^{j}_{n \, \beta \, \lambda \, K} \} \Psi
=
\{ \left( \overline{a}^{i \, \kappa}_{n \, \alpha \, H} \right)^{\dag} , a^{j}_{n \, \beta \, \lambda \, K} \} \Psi$$
and $$\{ a_{n \, \alpha \, \kappa \, H}^{i}, \left(a_{n \, \beta \, K}^{j \, \lambda}\right)^{\dag} \}
\Psi =
 2 \delta^{ij} \delta_{\kappa}^{\lambda} \, \delta_{\alpha \beta} \, \delta_{HK} \Psi.$$ Although
  the above
  operators transform covariantly\footnote{Connection terms do not appear when acted upon
  functions; when acted upon vectors curvature corrections appear.}; the expressions
  themselves are not covariant but attached to an inertial coordinate system just like the
  ordinary creation and annihilation operators are in quantum field theory.  We show now how
  the usual Fock basis emerges in the distributional limit; define $$a_{\mathbf{p} \, n \, \gamma \, \kappa \, H}^{T}
   = \frac{1}{\sqrt{2T}} \sum_{j : x^{\, j}_n \in \mathcal{P}^{T}_n} \left(
   \frac{1}{n} \right)^{2} e^{i p^{\alpha} \sum_{K} ( tr(x^{j \beta \, + \, 0}_{n \, K}) +
   tr({x^{j \beta \, +}_{n \, 0 \, K}}^{\dag}) ) \eta_{\alpha \beta}/2} a^{j}_{n \,
    \gamma \, \kappa \, H}$$ then direct computation gives
$$\{a_{\mathbf{p} \, n \, \gamma \, \kappa \, H}^{T}, a_{\mathbf{q} \, n \, \delta \, \lambda \, K}^{T} \}
= 0$$ and
$$\{a_{\mathbf{p} \, n \, \gamma \kappa \, H}^{T}, a_{\mathbf{q} \, n \, \delta \, K}^{T \, \lambda  \dag } \}
= $$ $$\frac{1}{2T} \sum_{j : x^{\, j}_n \in \mathcal{P}^{T}_n} 2 \left( \frac{1}{n} \right)^4 e^{i ( p^{\alpha} - q^{\alpha}) \sum_{S} ( tr(x^{j \beta \, + \, 0}_{n \, S}) + tr({x^{j \beta \, +}_{n \, 0 \, S}}^{\dag}))
 \eta_{\alpha \beta}/2} \delta_{\kappa}^{\lambda} \delta_{\gamma \delta} \, \delta_{HK}.$$
 The latter reduces to
  $$ (2 \pi)^3 \delta^{(3)}(\vec{p} - \vec{q}) 2 \delta_{\kappa}^{\lambda} \delta_{\gamma
  \delta} \delta_{HK}$$ in the limit $n \rightarrow \infty$ and this independently of $T$.
  Obviously, we take the limit $T \rightarrow \infty$ since we do not wish to violate space
   time translation invariance.  Free Fermi quantum field theory arises in this construction
   by atomizing non commutative space-time itself; indeed in the limit
    $n \rightarrow \infty$, the latter is partitioned by infinitesimal hypercubes which
    correspond to independent variables each. \\* \\ Local Fermi operators for the corresponding ''anti-particles'' 
   are given by $$b^{i \, (k) \dag}_{n \, \alpha \, \kappa \, H} =
   \widehat{\partial}_{x^{i \, (-) \, \alpha \, \kappa}_{n \, H}} -
    x^{i \, (-) \, \alpha \, \dag}_{n \, \kappa \, H}$$ and  $$b^{i \, (k) \, \kappa}_{n \, \alpha \, H}
     = \widehat{\partial}_{{x^{i \, (-) \, \alpha \, \dag}_{n \, \kappa \, H}}} -
      x^{i \, (-) \, \alpha \, \kappa}_{n \, H}$$ as well as their complex conjugates.  It is
      easily verified that
   $$ \{ b^{i \kappa}_{n \, \alpha \, H} , b^{j \, \lambda}_{n \, \beta \, K} \} = 0,$$
   $$ \{ b^{i \, \kappa}_{n \, \alpha \, H} , b^{j \, \dag}_{n \, \beta \, \lambda \, K} \} =
   - 2 \delta^{ij} \delta^{\kappa}_{\lambda} \delta_{\alpha \beta} \delta_{HK},$$
   $$ \{ b^{i \dag}_{n \, \alpha \, \kappa \, H} , a^{j}_{n \, \beta \, \lambda \, K} \} = 0$$ and
   $$ \{ b^{i \dag}_{n \, \alpha \, \kappa \, H}, a^{j \, \lambda \dag}_{n \, \beta \, K} \} = 0.$$  \\
   Given that the former anti-commutator algebra has the wrong sign, we have to
   redefine it so that it becomes positive definite (by defining the involution $b^s = - b^{\dag}$).  The particle field operator as well as the associated algebra, can now be constructed in the usual way in the quaternion formalism by using the abelian product in the anti-commutator.  The hidden complex polarizations of the uni-modal formalism are not accounted for in the QFT approximation, since as argued before, these are not measurable anyhow and should therefore not appear in a theory which is only concerned about measurable properties.  \\* \\
It remains to be seen whether the atomization procedure of quantum field theory is something
 real or merely a convenient way of approximating solutions to an operator valued non-commutative
  field theory on non-atomized, non-abelian space-time.  Indeed, since Bell's
 theorem goes down the drain, this might very well be the case given that the way quantum
  field theory is derived from a hypothetical non commutative operator valued field theory
   is exactly the same as thermodynamics follows from classical field theory.  We explore
    this line of thought further in the next paper.  \\* \\
In the next section, we examine the transformation groups of the kinematical structure,
assign particles/antiparticles to the Fermi operators and calculate the electromagnetic
 charges.
\subsection{Gauge groups of $\mathcal{M}$.}  Prior to addressing the question of dynamics, we have to determine the group of local coordinate transformations of our non commutative space-time.  The latter have to preserve the special structure of the coordinate systems at hand; therefore the maximal transformation group is generated by
\begin{itemize}
\item the local diffeomorphism group $\textrm{Diff}(\mathcal{S})$ in the coordinates $x^{\alpha}$,
\item local, structure preserving transformations changing the quaternion base over $C$, that is $SU(3)$,
\item local, structure preserving transformations changing the complex base over $R$, that is $U(1)$.
\end{itemize}
The first one constitutes the local coordinate transformations in the super structure,
 while the last two correspond to the unbroken symmetries of the standard model and by
definition, the super diffeomorphisms commute with the local gauge transformations.
It remains to be seen how the local $SU(2)$ gauge group as well as the Higgs mechanism
 arises from similar arguments as those above.  It is expected that the non commutative
 structure allows one to assign a small positive or imaginary effective mass -with respect to
 the abelian part of the coordinate system- to photons and gluons, but these should correspond
 to quantum or gravitational effects. \\* \\  First, let us assign the \emph{conventional} particle/anti-particle structures
 to particular coordinates, or better to the respective creation and annihilation operators.
For fixed $\alpha$ and using the language of the first Standard Model generation, $x^{\alpha \, (-) \, 0}_{H}$
 corresponds to an electron anti-neutrino, has electric charge $e = 0$ and transforms as a
singlet under $SU(3)$, $x^{\alpha \, (-) \, k}_{H}$ corresponds to a color $3$ multiplet associated
 to a down quark $d$ with $e = - 1/3$, $\overline{x^{\alpha \, (-) \, 0}}_{H}$ to an electron
$e = - 1$,  $\overline{x^{\alpha \, (-) \, k}}_{H}$ to a color $\overline{3}$ multiplet associated
 to a up anti-quark $\overline{u}$ with $e = - 2/3$.  $\dag$ maps the right (left) handed particles
 to their
respective right (left) handed anti-particles, so we have completed one generation of the extended standard model; that is, right as well as left handed neutrino's exist.   \\* \\
 We prove now the above remarks; the complex Euclidean quaternions are generated by
elements satisfying $$\left[q_{k} , q_{l} \right] = - \overline{i} (2 \epsilon_{klm}) q^{m \,
\dag}.$$  Now, it is most easily seen that the complex linear mappings of unit determinant
preserving this algebra are defined by $$q_k \rightarrow {U^{\dag \, l}_k} q_l$$ where
$\mathbf{U}$ is a $SU(3)$ transformation.  The reason why we wrote $$\left[ q_{k} , q_{l} \right]
= - \overline{i} (2 \epsilon_{klm}) q^{m \,\dag}$$ is because this expression is invariant under rotations in the complex plane which leave the quaternion basis invariant.  Therefore, it is important to distinguish the complex unit from the quaternion unit and only real combinations $\bar{y}x$ can be disgarded.  There is a residual freedom which modifies the complex conjugation by a factor of $e^{i \theta}$ and  $q_{k} \rightarrow e^{i \theta/3} q_{k}$; it is then easily verified that the latter preserves the complex quaternion algebra.  The $U(1)$ factor constructed above attributes the following $U(1)$ charges to the right handed standard model particles and anti particles $$0, 1/3, 1, 2/3, -1/3, -2/3, -1$$ and these turn out to be the correct numbers. \\* \\  Let us summarize what we have achieved so far:
\begin{itemize}
\item We have found five natural mathematical axioms from which we can derive (i) the number
 of generations of the standard model (ii) the correct number of particles and anti-particles per generation (iii) the bundle structure associated to the unbroken symmetry groups of the standard model.
\item We have explicitly shown that the statistical mechanics of a field theory with infinitesimal non-abelian dimensions coincides with free Fermi quantum field theory.
\item We have shown that abelian space times have no curvature.
\item The gauge charges drop out correctly, at least for the electromagnetic and strong
 interactions.
\end{itemize}
The least one can say is that these results are promising and the development of a full
dynamics as well as the construction of the necessary mathematical framework is postponed
to the follow-up paper.  It is important to note that our results indicate that the strong
and electromagnetic forces are essentially different from gravitation; the notable exception
for now being given by the weak interactions\footnote{We return to this issue in a follow up paper}. \\* \\
Finally, let me slightly speculate about the physical picture which emerges from this work :
 space-time turns out to be a non-commutative manifold and physical processes are locally
 causal in the non-abelian sense.  Quantum effects arise from physical
  operators penetrating the non abelian part of the algebra so that
   apparent causality violating effects, that is when one forgets about the non-abelian part
   of the metric, arise.
\section{Philosophical contemplations.}
Finally, let us consume the liberty to make some philosophical side remarks and indicate how
the dynamics should be constructed.  On the gauge theory side, there are no Yang Mills
action principles as elucidated in section $3.5$ since these generate unwanted vacuum
solutions, are not gauge invariant off-shell and therefore ignore self interactions as well as
 mass corrections.  Gauge theory is nothing but a non-linear self coupling mediated by the gravitational field (the retarded
Green's function).  The gauge bosons are therefore constructed from and acting upon fermionic
\emph{operators} defined upon the natural tangent bundle of non-abelian space-time.  Any maximal
set of such operators spans the three Standard Model generations as well as a fourth, exotic,
timelike generation.  A suitable interpretation for the latter still has to be conceived.  The
 general idea on the gravity side is that the groupage of such operators in an operator valued
  vierbein span the (operational) metric (hence, invertibility is a \emph{demand}).  Generalized local Lorentz covariance is abandonned for
  several reasons (a) we want to improve gravity by demanding that matter \emph{is} geometry (b) the theory at hand
  should be able to dynamically generate space and time and in our view, space is matter (just as time is some exotic form of it).  Hence, gravity is based upon
  the Weitzenbock connection, that is, I expect it to be a pure torsion theory with vanishing curvature.  There would be no cosmological constant, since
  (a) mass is generated dynamically and (b) there is no quantum (in the sense explained below) mechanical vacuum energy.  \\* \\
Quantum mechanics is in my view an effective bookkeeping arising from simulating the non-abelian
effects generated by extended space-time in an abelian setting.  One certainly would have to search for a weak form of the superposition principle; indeed,
what quantum physicists call the superpostion of two states, could be nothing but two weakly interacting, essentially
four dimensional over $R$, \emph{real} brane like solutions moving around in extended
space-time.  This weak interaction would be crucial since at some point it would have to become strong in order to
trigger observation.  It is obvious that at this point, these things are mere speculation,
but they are genuine possibilities residing within the kinematical structure at hand and constitute a ''picture'' which solves the measurement problem.
\\* \\  Finally, let me say something about the initial value problem; it appears to me that within the language
of differential equations, only one point of view is sensible.  I call it the chaotic universe:
it is a universe such that for any initial value problem, a slicing can be found such that the latter
can be mapped into it with arbitrary precision.  If the world is truely four dimensional,
then we simply are somewhere in it, but trying to point out this ''where'' \emph{a priori} is in direct conflict with the four
dimensional character itself.

\end{document}